\documentclass[twocolumn]{aa} 
\usepackage{graphicx}
\usepackage[final]{hyperref}
\usepackage{rotating}
\usepackage{float}
\usepackage{txfonts}
\usepackage{float}
\usepackage{siunitx}
\usepackage{tabularx}
\usepackage{color,hyperref}
\usepackage{geometry,fancyhdr}
\usepackage{multicol,multirow}
\usepackage{arydshln}
\usepackage{caption}
\usepackage{subcaption}
\usepackage{lscape}
\usepackage[flushleft]{threeparttable}
\hypersetup{colorlinks,breaklinks,
            linkcolor=blue,urlcolor=blue,
            anchorcolor=blue,citecolor=blue}
\usepackage{caption}
\usepackage{longtable}
\usepackage{comment}
\bibpunct{(}{)}{;}{a}{}{,} 
\usepackage{txfonts}
\begin{document} 

   \title{The Effelsberg survey of FU~Orionis and EX~Lupi objects II.}
    \subtitle{H$_2$O maser observations}

   \author{Zs. M. Szabó\inst{1,2,3,4,\thanks{Member of the International Max Planck Research School (IMPRS) for Astronomy and Astrophysics at the Universities of Bonn and Cologne.}},
          Y. Gong\inst{1},
          W. Yang\inst{1},
          K. M. Menten\inst{1},
          O. S. Bayandina\inst{5},
          C. J. Cyganowski\inst{2},
          Á. Kóspál\inst{3,4,6,7},
          P. Ábrahám\inst{3,4,6},
          A. Belloche\inst{1}
          \and
          F. Wyrowski\inst{1}
          }
    \institute{Max-Planck-Institut für Radioastronomie, Auf dem Hügel 69, 53121 Bonn, Germany\\
              \email{zszabo@mpifr-bonn.mpg.de}
    \and
    Scottish Universities Physics Alliance (SUPA), School of Physics and Astronomy, University of St Andrews, North Haugh, St Andrews, KY16 9SS, UK
    \and
    Konkoly Observatory, Research Centre for Astronomy and Earth Sciences, E\"otv\"os Lor\'and Research Network (ELKH), Konkoly-Thege Mikl\'os \'ut 15-17, 1121 Budapest, Hungary
    \and
    CSFK, MTA Centre of Excellence, Budapest, Konkoly Thege Miklós út 15-17., H-1121, Hungary
    \and
    INAF - Osservatorio Astrofisico di Arcetri, Largo E. Fermi 5, 50125 Firenze, Italy
    \and
    ELTE E\"otv\"os Lor\'and University, Institute of Physics, P\'azm\'any P\'eter s\'et\'any 1/A, H-1117 Budapest, Hungary
    \and
    Max-Planck-Institut für Astronomie, K\"onigstuhl 17, D-69117 Heidelberg, Germany
    }
    
   \date{Received ; accepted}

 
  \abstract
   {FU~Orionis (FUor) and EX~Lupi (EXor) type objects are two groups of peculiar and rare pre-main sequence low-mass stars that are undergoing powerful accretion outbursts during their early stellar evolution.  Though water masers are widespread in star forming regions and are powerful probes of mass accretion and ejection on small scales, little is known about the prevalence of water masers toward FUors and EXors.}
   {We aim to perform the first systematic search for the 22.2\,GHz water maser line in FUors and EXors in order to determine its overall incidence in these eruptive variables and facilitate follow-up high angular resolution observations.}
   {We used the Effelsberg\,100-m radio telescope to observe the H$_2$O ($6_{16}-5_{23}$) transition at 22.2\,GHz toward a sample of 51 eruptive young stellar objects.}
   {We detect 5 water masers in our survey; 3 are associated with eruptive stars, equivalent to a detection rate of $\sim$6\% for our sample of eruptive sources. These detections include one EXor, V512~Per (also known as SVS~13 or SVS~13A), and two FUors, Z~CMa and HH~354~IRS.  This is the first reported detection of water maser emission towards HH~354~IRS. 
   We also detect water maser emission in our pointing towards the FUor binary RNO~1B/1C, which most likely originates from the nearby deeply embedded source IRAS~00338$+$6312 ($\sim$4\arcsec\, from RNO~1B/1C). Emission was also detected from H$_2$O(B) (also known as SVS~13C), a Class\,{\footnotesize 0} source $\sim$30\arcsec\,from the EXor V512~Per.  The peak flux density of H$_2$O(B) in our observations, 498.7\,Jy, is the highest observed to date, indicating that we have serendipitously detected a water maser flare in this source.
   In addition to the two non-eruptive Class\,{\footnotesize 0} sources (IRAS 00338$+$6312 and H$_2$O(B)/SVS~13C), we detect maser emission towards  one Class\,{\footnotesize 0/I} (HH~354~IRS) and two Class\,{\footnotesize I} (V512~Per/SVS~13A and Z~CMa) eruptive stars.}
   {Despite the low detection rate, we demonstrate the presence of 22.2\,GHz water maser emission in both FUor and EXor systems, opening the way to radio interferometric observations to study the environments of these eruptive stars on small scales. Comparing our data with historical observations spanning several decades suggests that multiple water maser flares have occurred in both V512~Per and H$_{2}$O(B).}
   \titlerunning{H$_2$O maser  detection towards FUors/EXors}
   \authorrunning{Szab\'o et al.}
   \maketitle
%
\section{Introduction} 
\label{sec:intro}
Low-mass young stellar objects (YSOs) are stars in the early stages of stellar evolution, specifically protostars and pre-main sequence (PMS) stars, which can undergo accretion-driven episodic outbursts.
Studies of outbursting objects provide crucial information on the formation and the evolution of Sun-like stars.
Amongst PMS stars, there are two small, but rather spectacular classes of outbursting low-mass YSOs: FU~Orionis and EX~Lupi-type stars (FUors and EXors for short, respectively). Members of both classes show major increases in their optical and near-infrared (NIR) brightnesses. FUors can brighten by up to 5 -- 6 magnitudes in the optical, triggered by enhanced accretion from the accretion disk onto the protostar \citep{HartmannKenyon1996, herbig1989_eso}.
This phase can last for several decades, or even centuries \citep[e.g.\ the recent review by][and references therein]{fischer2022}.
For example, the prototype of the FUor class, FU~Orionis, went into outburst in 1936 \citep{wachmann1954}, and remains in a highly active state. 
After a few other objects were observed to experience similar outbursts, \citet{herbig1977} defined the FUor class, which continues to increase in size as new FUor-type objects are identified  
\citep[e.g.,][]{audard2014,szegedi-elek2020} and currently contains more than a dozen objects.
The EXor class was defined by \citet{herbig1989_eso}, based on the properties of the prototype star EX~Lupi, and currently also includes more than a dozen objects \citep[e.g.,][]{audard2014,park2022}. EXors can brighten by up to 1 -- 5 magnitudes in the optical and remain in a bright state for a few months or a few years \citep[see e.g.,][]{sepic2018}; furthermore, their outbursts are recurring \citep[e.g.,][]{audard2014,cruz-saenz2022}.

Interstellar masers are powerful tools for studying the physics of star formation on small scales, frequently probing regions of enhanced density and temperature \citep[e.g.,][]{1992ARA&A..30...75E,2014ARA&A..52..339R}. While masers have been substantially used to probe both low- and high-mass star formation regions \citep[e.g.,][]{abraham1981,omodaka1999,hirota2011,furuya2001,furuya2003}, so far little information exists on masers in FUors/EXors. Pioneering studies found compact maser emission in the 1720\,MHz hyperfine structure line of hydroxyl (OH) toward the archetypal FUor V1057~Cyg \citep{1973ApJ...185L..71L}. This emission, which comes from the immediate vicinity of the star \citep{1974ApJ...190L.125L} and is highly time variable \citep{1981A&A....93...79W}, is unique in the literature. 
The 22.2\,GHz transition of water (H$_2$O) is the most widespread interstellar maser \citep[see, e.g.,][and references therein]{gray2012}. It has been detected towards numerous low- to high-mass star forming regions in the Milky Way \citep[see e.g.][]{2022AJ....163..124L}. Pumping models indicate that 22.2\,GHz water masers are excited at elevated temperatures ($\sim$500\,K) and densities ($10^{8-9}$\,cm$^{-3}$), which are typically found in the compressed post-shock regions of jets/outflows from YSOs \citep{elitzur1989a,elitzur1989b,gray2012,gray2022}. 
With very-long-baseline interferometry (VLBI), multi-epoch observations of water masers associated with protostellar outflows can be used to study mass accretion and ejection \citep[see, for example,][]{2016MNRAS.460..283B,2019A&A...631A..74M}.
This suggests that water masers could potentially serve as valuable probes of mass accretion and ejection in FUors/EXors. 

Despite the fact that water masers are closely associated with mass accretion and ejection in protostars, a systematic search for 22.2\,GHz H$_2$O masers in FUors/EXors has not yet been performed.
Hence, the overall incidence of 22.2\,GHz water masers in these classes of eruptive objects is unknown. 
In this paper, we present the first dedicated 22.2\,GHz water maser survey of low-mass young eruptive stars,  using the Effelsberg~100-m telescope.  Our single-dish survey is a first step in investigating water masers in low-mass outbursting systems, aimed at investigating the existence and prevalence of water masers in these objects and identifying targets for follow-up interferometric observations.
This paper is the second in a series \citep[the first being][]{szabo2023} presenting radio and (sub)millimeter observations of FUors and EXors and their natal environments, and is organized as follows. 
In Sect.~\ref{sec:observations}, we summarize our observations. In Sect.~\ref{sec:res}, we present our results, focusing on sources with water maser detections. In Sect.~\ref{discussion}, we discuss our results, and in Sect.~\ref{sec:conclusions} we summarize our most important findings. 

\section{Observations} 
\label{sec:observations}

The H$_2$O $J_{K_a,K_c} = $ $6_{16}-5_{23}$ transition (rest frequency 22235.0798\,MHz, from the JPL Molecular Spectroscopy database\footnote{\url{https://spec.jpl.nasa.gov/}},  \citealt{1998JQSRT..60..883P}) was observed simultaneously with the three lowest metastable NH$_{3}$ transitions ($(J,K) = (1,1), (2,2)$ and $(3,3)$), which were presented in Paper\,{\footnotesize I} \citep{szabo2023}. The observations were carried out on 2021 November 18,
November 23, and 2022 January 25 using the Effelsberg~100-m telescope in Germany\footnote{The 100-m telescope in Effelsberg is operated by the Max-Planck- Institut für Radioastronomie (MPIfR) on behalf of the Max-Planck Gesellschaft (MPG).} (project id: 95-21, PI: Szab\'o). 
The sample consisted of 51 sources: 33 FUors, 13 EXors, and 5 Gaia alerts. Gaia alert sources were chosen from the variable sources identified by the Gaia Photometric Science Alerts system \citep[][]{hodgkin2021} based on light curve characteristics and luminosities similar to those of FUors/EXors. 
Five Gaia alert sources in our sample are yet to be classified; one source, Gaia18dvy, is listed with its Gaia alert name (Table~\ref{tab:appendix_non-detections}) but counted as a FUor based on its classification by \citet{szegedi-elek2020}.

Our observations were performed in position-switching mode with an off-position at an offset of 5\arcmin\, east of our targets in azimuth.  
During our observations, the 1.3\,cm double beam and dual polarization secondary focus receiver was employed as the frontend, while the Fast Fourier Transform Spectrometers (FFTSs) were used as the backend. Each FFTS provides a bandwidth of 300\,MHz and 65536 channels, which gives a channel width of 4.6\,kHz, corresponding to a velocity spacing of 0.06\,km\,s$^{-1}$ at 22.2\,GHz. The actual spectral resolution is coarser by a factor of 1.16 \citep{2012A&A...542L...3K}. 

At the beginning of each observing session, pointing and focus were verified towards NGC~7027. On 2021 November 18 we also targeted W75N, known for its H$_2$O and NH$_3$ emission, to make sure that the system was working properly (see Appendix\,\ref{sec:w75n}). Pointing was regularly checked on nearby continuum sources, and was found to be accurate to about 5$\arcsec$. NGC~7027 was also used as our flux calibrator, assuming a flux density of $\sim$5.6\,Jy at 22.2\,GHz \citep{ott1994}.
The on-source integration time was 2.5 minutes per spectrum, and during each observing epoch, 4 spectra per source were obtained. 

The majority of our sources were observed on 2021 November 18 and 23 (see Tables~\ref{tab:fit_results} and \ref{tab:appendix_non-detections}).
On 2021 November 18, we detected H$_{2}$O maser emission toward V512~Per (SVS~13A), RNO~1B/1C, and HH~354~IRS. To study the time variability of the maser emission, we re-observed detected sources in as many subsequent epochs as possible (see Table~\ref{tab:fit_results}), within the constraints of our allocated observing sessions.   
For Z~CMa, which was known to have water maser emission \citep[][]{moscadelli2006} but could not be observed in November 2021 due to time constraints, we searched for short-term maser variability by observing this source for two 4$\times$2.5 minute blocks separated by 2.5\,hours in January 2022. 
No variability was detected on this timescale, so all 8 spectra of Z~CMa were averaged for the subsequent analysis.
We note, that due to the weak detection of the water maser in HH~354~IRS, the spectrum was spectrally smoothed by a factor of 2 using the \texttt{smooth} built-in function in CLASS. The smoothed spectrum is presented throughout this paper.
Having detected unusually high-amplitude (factor of $\sim$4 with respect to the previous observation) and rapid variability in the H$_2$O maser spectra towards V512~Per (SVS~13A) (see Sect.~\ref{sec:res_v512per}), we also carried out nine-point observations and 1\arcmin$\times$1\arcmin\ On-The-Fly (OTF) mapping of this source on 2022 February 5 to investigate whether emission from nearby sources in the telescope sidelobes could be contributing to the observed emission. Consequently, we serendipitously detected strong water maser emission toward H$_2$O(B) (SVS~13C), which is 30\arcsec\, from V512~Per (SVS~13A) (see Sect.~\ref{sec:res_v512per} and \ref{sect:h2ob-followup}). We also performed single-pointing observations towards H$_2$O(B) during this epoch.

We adopted the method introduced by \citet{2012A&A...540A.140W} for our spectral calibration which resulted in a calibration uncertainty of about 15\%. The half-power beam width (HPBW) was about 40\arcsec\,at 22\,GHz and the main beam efficiency was 60.2\% at 22\,GHz. The conversion factor from flux density, $S_{\nu}$, to main beam brightness temperature, T$_{\rm mb}$, was T$_{\rm mb}/\rm S_{\nu}$=1.73\,K/Jy. Typical RMS noise levels for observations of detected sources are given in Table~\ref{tab:fit_results} and 3$\sigma$ upper limits for non-detections are given in Table~\ref{tab:appendix_non-detections}.
 
The data were reduced using the GILDAS/CLASS package developed by the Institut de Radioastronomie Millim{\' e}trique (IRAM) \footnote{\url{https://www.iram.fr/IRAMFR/GILDAS/}} \citep{2005sf2a.conf..721P,2013ascl.soft05010G}. 
For each target, spectra observed on the same day were averaged to improve the signal-to-noise ratio prior to subtracting a linear baseline. 
Velocities are presented with respect to the local standard of rest (LSR) throughout this paper.
 
\section{Results} 
\label{sec:res}
Of our 51 targets, we detected $>$3$\sigma$ water maser emission towards two FUors (Z~CMa and HH~354~IRS) and one EXor (V512~Per/SVS~13A), corresponding to a detection rate of $\sim$6\% towards eruptive stars. 
We also serendipitously detected water maser emission towards two non-eruptive embedded protostars, which we discuss in Sects.~\ref{sect:h2ob-followup} and \ref{sec:fuors-rno1b1c}. 
The basic parameters of sources with maser detections, including types, coordinates, distances, and evolutionary classifications are listed in Table~\ref{tab:coordinates}. In all, we detected water masers in two non-eruptive Class\,{\footnotesize 0} sources (IRAS~00338$+$6312 and H2O(B)/SVS 13C) and in one Class\,{\footnotesize 0/I} (HH~354~IRS) and two Class\,{\footnotesize I} (V512 Per/SVS 13A and Z CMa) eruptive objects, using the standard classification scheme \citep[see, e.g.,][]{greene1994,evans2009}.

For sources with water maser detections, we fitted each velocity component with a Gaussian to obtain its LSR velocity ($\varv_{\rm LSR}$), line width ($\Delta \varv$), and peak flux density ($S_{\nu}$), given in Table~\ref{tab:fit_results}. The peak flux densities of detected water masers vary from 0.11\,Jy to 498.7\,Jy, spanning over 3 orders of magnitude.  The observed maser velocities are within 10\,km\,s$^{-1}$ of the systemic cloud velocities measured from NH$_{3}$ emission. 
While shock velocities of $\gtrsim$50\,km\,s$^{-1}$ are expected in theoretical models \citep[e.g.,][]{1989ApJ...346..983E}, the modest velocity offsets between water masers and dense gas observed in our sample are generally consistent with observations of water masers towards high-mass YSOs \citep[e.g.,][Fig.\ 4 and Fig.\ 16 respectively]{2009A&A...507..795U,cyganowski2013}.
Isotropic H$_2$O maser luminosities, $L_{\rm H_{2}O}$, were calculated as \citep[e.g.,][]{anglada1996,urquhart2011,cyganowski2013}:
\begin{equation}\label{f.lum}
    \left[\frac{L_{\rm H_{2}O}}{L_{\odot}}\right]=2.3\times 10^{-8} \left[\frac{\int S_{\nu}{\rm d}\varv}{\rm Jy\;km \;s^{-1}}\right] \left[\frac{D}{\rm kpc}\right]^{2}\;,
\end{equation}
where $D$ is the distance to the target (see Table~\ref{tab:coordinates}). Estimating the isotropic H$_2$O maser luminosities of individual velocity components separately, we find a range of $L_{\rm H_{2}O}$ of 7.9$\times 10^{-10} L_{\odot}$ to 6.1$\times 10^{-7}L_{\odot}$ (see Table~\ref{tab:fit_results}). 

\begin{table*}[!htbp]
\small
\caption{Basic information about the sources towards which water maser emission was detected at 22.2\,GHz.}   
\label{tab:coordinates}      
\centering                                      
\begin{tabular}{ccccccc}        
\hline\hline                        
\multirow{2}{*}{Name}       & Type       & R.A. (J2000)                    & Dec. (J2000) & Distance  & \multirow{2}{*}{Classification} &\multirow{2}{*}{Reference} \\ 
    & FUor/EXor & ($^{\rm h}$ $^{\rm m}$ $^{\rm s}$)   & ($^{\circ}$ $\arcmin$ $\arcsec$) & (pc) & & \\
\hline \hline      
Z~CMa       & FUor & 07 03 43.15 & $-$11 33 06.2 & 1125 & Class\,{\footnotesize I} & 1, 2, 3 \\
HH~354~IRS		& FUor & 22 06 50.37 & +59 02 45.9 & 750 & Class\,{\footnotesize 0/I} & 1, 4, 5, 6 \\
V512~Per (SVS~13A)	& EXor & 03 29 03.75 & +31 16 03.9 & 275 & Class\,{\footnotesize I} & 1, 7, 8 \\
\hdashline
IRAS~00338$+$6312 (close to RNO~1B/1C)   & -- & 00 36 46.30 & +63 28 54.0 & 965 & Class\,{\footnotesize 0} & 7, 9 \\
H$_2$O(B) (SVS~13C)$^*$ & -- & 03 29 01.35 & 31 15 40.4 & 275  & Class\,{\footnotesize 0} &  7, 10 \\
\hline \hline
\end{tabular}
\flushleft
\tablefoot{$^*$The coordinates given for H$_2$O(B) were derived from the OTF mapping. \\ 
References for distance and classification: 1 -- \citet{audard2014}; 2 -- \citet{dong2022}; 3 -- \citet{gramajo2014}; 
4 -- \citet{reipurth1997b}; 5 -- \citet{bronfman1996}; 6 -- \citet{reipurth1997b}; 7 -- \citet{bailerjones-edr3}; 8 -- \citet{diaz-rodriguez2022}; 9 --\citet{quanz2007}; 10-- \citet{plunkett2013}.}
\end{table*}

In the following subsections, we discuss our results for sources with detected water masers.
Our non-detections are presented in Appendix~\ref{sec:app_non-detections}, where
Table~\ref{tab:appendix_non-detections} lists the targeted sources along with their types, coordinates, 3$\sigma$ upper limits, whether they were previously searched for 22.2\,GHz maser emission and if so
the reference, the date of observation in the current survey,
their classification and reference, and distances.

For 31 sources in our sample, no previous  observations of the 22.2\,GHz water maser line have been reported in the literature.

\begin{table*}[!htbp]
\small
\caption{Properties of observed water maser features.}  
\label{tab:fit_results}      
\centering                                      
\begin{tabular}{lcrrrrlr}    
\hline\hline          
\multirow{3}{*}{Source} & \multirow{3}{*}{Date} & \multicolumn{5}{c}{H$_2$O}  & \multicolumn{1}{c}{NH$_3$ (1,1)}  \\
\cline{3-7} 
\cdashline{8-8}
&       & \multicolumn{1}{c}{$\varv_{\rm LSR}$}  &  \multicolumn{1}{c}{$\Delta \varv$}        & \multicolumn{1}{c}{RMS} & \multicolumn{1}{c}{$S_{\nu}$}  &  \multicolumn{1}{c}{$L_{\rm H_2O}$} & \multicolumn{1}{c}{$\varv_{\rm LSR}$}  \\
& yyyy-mm-dd & \multicolumn{1}{c}{(km\,s$^{-1}$)} & \multicolumn{1}{c}{(km\,s$^{-1}$)}   & \multicolumn{1}{c}{(Jy)} & \multicolumn{1}{c}{(Jy)}   & \multicolumn{1}{c}{($L_\odot$)}     & \multicolumn{1}{c}{(km s$^{-1}$)} \\
\hline \hline                                 
V512~Per (SVS~13A)$^*$  & 2021-11-18 &  $6.31$ $(0.06)$  & $1.20$ $(0.06)$ & $0.07$ & $20.20$ $(0.07)$ & 3.7$\times10^{-8}$ & \multirow{5}{*}{$8.45$ $(0.01)$}  \\
V512~Per$^*$  & 2021-11-18 &  $8.46$ $(0.06)$  & $0.86$ $(0.06)$ & $0.07$ & $19.24$ $(0.07)$ & 3$\times10^{-8}$ & \\
V512~Per$^*$ & 2021-11-18 &  $10.91$ $(0.06)$  & $0.74$ $(0.06)$ & $0.07$ & $19.02$ $(0.07)$ & 2.4$\times10^{-8}$  &  \\
V512~Per$^*$ & 2021-11-18 &  $11.68$ $(0.06)$  & $0.56$ $(0.06)$ & $0.07$ & $16.88$ $(0.07)$ & 1.6$\times10^{-8}$  &  \\
V512~Per$^*$ & 2021-11-18 &  $13.23$ $(0.06)$  & $1.56$ $(0.06)$ & $0.07$ & $1.72$ $(0.07)$ & 3.6$\times10^{-9}$ &  \\
\cdashline{1-8}
V512~Per$^*$ & 2021-11-23 &  $6.30$ $(0.03)$   & $1.08$ $(0.01)$  & $0.04$ & $55.81$ $(0.05)$ & 9.7$\times10^{-8}$ & \multirow{5}{*}{$8.45$ $(0.01)$}   \\
V512~Per$^*$ & 2021-11-23 &  $8.42$ $(0.03)$   & $0.95$ $(0.01)$  & $0.04$ & $86.38$ $(0.05)$ & 1.5$\times10^{-7}$ &   \\
V512~Per$^*$ & 2021-11-23 &  $10.95$ $(0.03)$   & $0.81$ $(0.01)$  & $0.04$ & $16.14$ $(0.05)$ & 2.8$\times10^{-8}$ &    \\
V512~Per$^*$ & 2021-11-23 &  $11.11$ $(0.03)$  & $1.28$ $(0.01)$ & $0.04$ & $15.85$ $(0.05)$ & 3.1$\times10^{-8}$ &   \\ 
V512~Per$^*$ & 2021-11-23 &  $13.40$ $(0.03)$  & $0.88$ $(0.01)$ & $0.04$ & $2.91$ $(0.05)$ & 4.1$\times10^{-9}$ &   \\
\cdashline{1-8}
V512~Per$^{*}$ & 2022-02-05 & $5.15$ $(0.01)$ & $0.74$ $(0.03)$ & $0.06$ & $2.09$ $(0.01)$ & 2.8$\times10^{-9}$ & \multirow{6}{*}{$8.45$ $(0.01)$} \\
V512~Per$^{*}$ & 2022-02-05 & $6.13$ $(0.01)$ & $0.67$ $(0.01)$ & $0.06$ & $21.31$ $(0.01)$ & 2.7$\times10^{-8}$ &  \\
V512~Per$^{*}$ & 2022-02-05 & $7.21$ $(0.01)$ & $0.69$ $(0.01)$ & $0.06$ & $4.33$ $(0.01)$ & 5.3$\times10^{-9}$ &  \\
V512~Per$^{*}$ & 2022-02-05 & $8.63$ $(0.01)$ & $0.97$ $(0.01)$ & $0.06$ & $7.99$ $(0.01)$ & 1.4$\times10^{-8}$ &  \\
V512~Per$^{*}$ & 2022-02-05 & $9.64$ $(0.01)$ & $1.12$ $(0.01)$ & $0.06$ & $5.81$ $(0.01)$ & 1.2$\times10^{-8}$ &  \\
V512~Per$^{*}$  & 2022-02-05 &   $11.82$ $(0.01)$ & $1.11$ $(0.01)$ & $0.06$ & $2.49$ $(0.05)$  & 4.5$\times10^{-9}$ &  \\ 
\cdashline{1-8}
H$_2$O(B) (SVS~13C)$^{**}$  & 2022-02-05 &  $5.07$ $(0.01)$ & $0.63$ $(0.01)$ & $0.06$ & $41.45$ $(0.05)$  & 4.8$\times10^{-8}$ & \multirow{5}{*}{$8.33$ $(0.01)$} \\ 
H$_2$O(B)$^{**}$  & 2022-02-05 &  $6.14$ $(0.01)$ & $0.66$ $(0.01)$ & $0.06$ & $498.7$ $(0.05)$ & 6.1$\times10^{-7}$ &  \\ 
H$_2$O(B)$^{**}$  & 2022-02-05 &  $7.22$ $(0.01)$ & $0.76$ $(0.01)$ & $0.06$ & $98.30$ $(0.05)$  & 1.3$\times10^{-7}$ & \\ 
H$_2$O(B)$^{**}$  & 2022-02-05 &  $8.62$ $(0.01)$ & $0.92$ $(0.01)$ & $0.06$ & $182.34$ $(0.05)$ & 3.1$\times10^{-7}$ &  \\ 
H$_2$O(B)$^{**}$  & 2022-02-05 &  $9.61$ $(0.01)$ & $0.96$ $(0.01)$ & $0.06$ & $126.73$ $(0.05)$ & 2.3$\times10^{-7}$ &  \\ 
\hline 
IRAS~00338$+$6312  & 2021-11-18  &  $-28.78$ $(0.01)$ & $1.72$ $(0.01)$ & $0.05$ & $2.61$ $(0.09)$ & 1.1$\times10^{-7}$   & \multirow{6}{*}{$-17.90$ $(0.06)$} \\
IRAS~00338$+$6312  & 2021-11-18  &  $-15.79$ $(0.05)$ & $0.96$ $(0.05)$ & $0.05$ & $0.37$ $(0.09)$  & 6.2$\times10^{-9}$  \\
IRAS~00338$+$6312  & 2021-11-23  &  $-28.78$ $(0.01)$ & $1.70$ $(0.01)$ & $0.04$ & $2.75$ $(0.09)$  & 1.1$\times10^{-7}$   \\
IRAS~00338$+$6312  & 2021-11-23  &  $-15.88$ $(0.02)$ & $0.69$ $(0.05)$ & $0.04$ & $0.39$ $(0.09)$  & 5.6$\times10^{-9}$  \\
IRAS~00338$+$6312  & 2022-01-25  &  $-28.48$ $(0.01)$ & $1.75$ $(0.02)$ & $0.04$ & $1.10$ $(0.09)$  & 4.1$\times10^{-8}$  \\
IRAS~00338$+$6312  & 2022-02-05  &  $-28.43$ $(0.01)$ & $1.69$ $(0.02)$ & $0.05$ & $1.63$ $(0.09)$  & 6.3$\times10^{-8}$  \\
\hline
HH~354~IRS& 2021-11-18  &  $1.18$ $(0.06)$    & $0.81$ $(0.10)$ & $0.07$ & $0.18$ $(0.03)$ & 2.1$\times10^{-9}$   & \multirow{3}{*}{$-1.52$ $(0.01)$} \\
HH~354~IRS& 2022-01-25  &  $-10.51$ $(0.02)$  & $0.54$ $(0.05)$ & $0.02$ & $0.11$ $(0.02)$  & 7.9$\times10^{-10}$   &  \\
HH~354~IRS& 2022-01-25  &  $5.04$ $(0.02)$    & $0.63$ $(0.05)$ & $0.02$ & $0.18$ $(0.02)$  & 1.6$\times10^{-9}$ &  \\
\hline
Z~CMa & 2022-01-25  &  $7.82$ $(0.01)$ & $1.36$ $(0.01)$   & $0.01$ & $2.36$ $(0.01)$ & 1.1$\times10^{-7}$ & $13.8$ $(0.02)$ \\
\hline \hline
\end{tabular}
\flushleft
\tablefoot{$^*$ From pointed observations towards V512~Per; possible contamination by H$_2$O(B) is discussed in Sect.~\ref{sect:h2ob-followup}.
$^{**}$From pointed observations towards H$_2$O(B) (see also Sect.~\ref{sect:h2ob-followup}).}
\end{table*}

\subsection{FUors}
\label{sec:fuors}

\subsubsection{Z~CMa}
\label{sec:zcma}
\begin{figure}[h!]
\centering 
\includegraphics[width=\columnwidth]{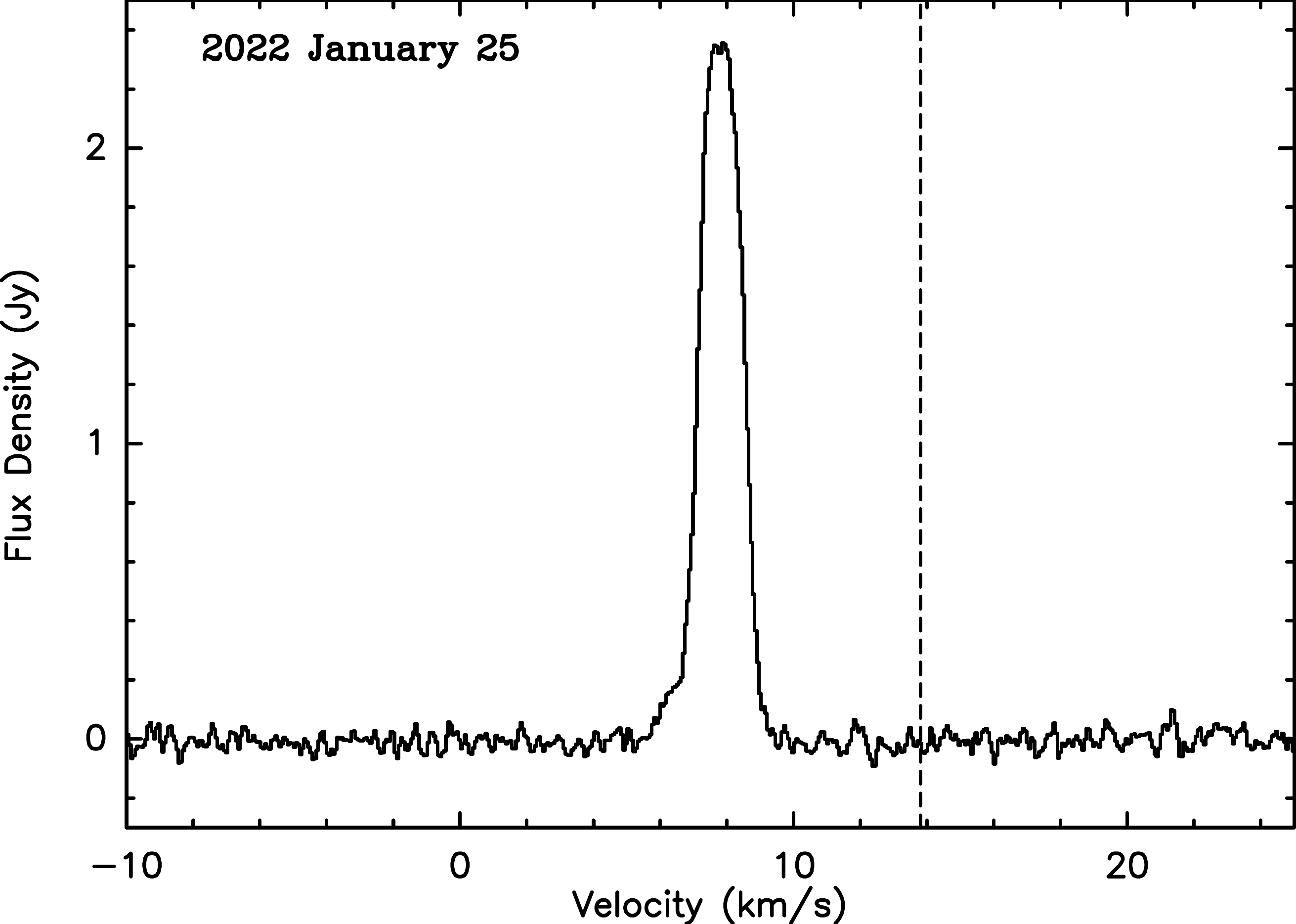}
\caption{H$_2$O maser spectrum of Z~CMa observed on 2022 January 25. The dashed vertical line indicates the $\varv_{\rm LSR}$ of 13.8\,km\,s$^{-1}$ from the NH$_3$ (1,1) transition \citep{szabo2023}.}
\label{fig:zcma}
\end{figure}

Z~CMa consists of an FUor (southwest component) and a Herbig~Ae/Be star (northeast component) that are only 0.1\arcsec\,apart \citep[][]{koresko1991,bonnefoy2017}.
Figure~\ref{fig:zcma} shows the H$_2$O maser spectrum observed toward Z~CMa, the only source among those detected observed at only one epoch (Sect.~\ref{sec:observations}).
As shown in Figure~\ref{fig:zcma}, there is only one bright maser feature, at $\varv_{\rm LSR}=$7.82\,km\,s$^{-1}$, blueshifted by $\sim$6\,km\,s$^{-1}$ with respect to the thermal NH$_3$ emission.
Although Z~CMa has been observed in many previous water maser studies \citep[][]{blitz1979,thum1981,deguchi1989,scappini1991,palla1993,moscadelli2006,sunada2007,bae2011,kim2018}, maser emission was detected only on 2003 March 14 \citep{moscadelli2006}, with a flux density of $\sim$2\,Jy at $\varv_{\rm LSR}=$ 14.3\,km\,s$^{-1}$. The maser component at $\varv_{\rm LSR}=$ 7.82\,km\,s$^{-1}$ is reported here for the first time.

\subsubsection{HH~354~IRS}
\label{sec:fuors-hh354}

HH~354~IRS, also known as IRAS~22051$+$5848 and L1165-SMM1 \citep[e.g.,][]{visser2002}, was classified as a FUor based on its CO first-overtone bandhead feature at 
$\sim$2.3\,$\mu$m \citep{reipurth1997a,connelley2018}.
This FUor was searched for H$_2$O maser emission multiple times between 1985 and 2005 \citep[][]{wouterloot1993,persi1994,sunada2007}, but no maser emission was reported. 

\begin{figure}[!htbp]
\includegraphics[width=\columnwidth]{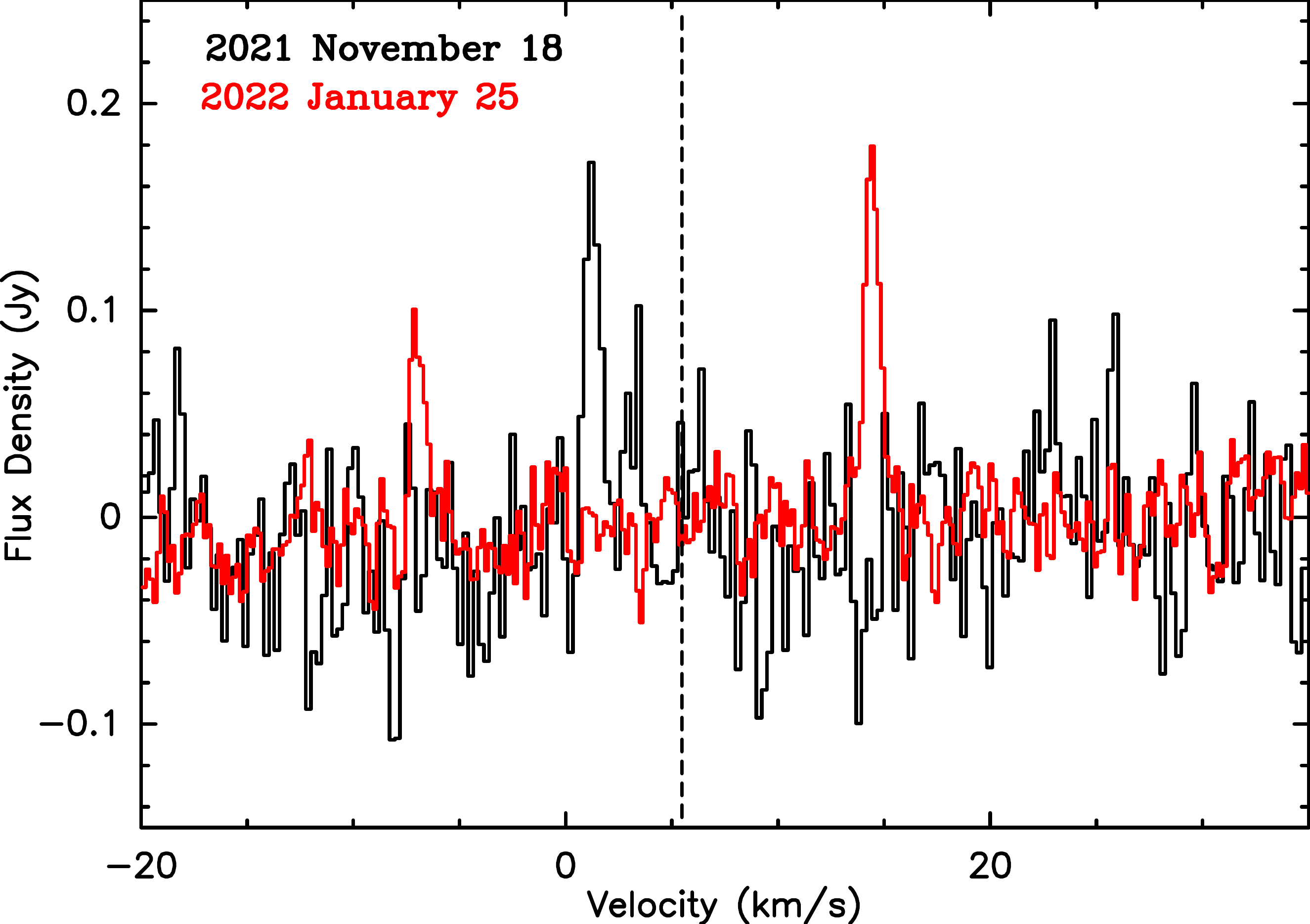}
\caption{Spectrally smoothed line profile of H$_2$O maser emission observed in HH~354~IRS on 2021 November 18 and 2022 January 25. The dashed vertical line indicates the average $\varv_{\rm LSR}$ of $-$1.47\,km\,s$^{-1}$ derived from the NH$_3$ (1,1) and (2,2) transitions \citep{szabo2023}.}
\label{fig:hh354}
\end{figure}

As shown in Figure~\ref{fig:hh354}, we detected weak H$_2$O maser emission (peak flux densities $<$0.2\,Jy, Table~\ref{tab:fit_results}) towards HH~354~IRS in two epochs.  These are the first detections of water maser emission towards this source. 
On 2021 November 18, we detected a weak H$_2$O maser at $\varv_{\rm LSR}=$1.18\,km\,s$^{-1}$. On 2022 January 25 we detected two features at $\varv_{\rm LSR}=$ $-$10.51 and $\varv_{\rm LSR}=$ 5.04\,km\,s$^{-1}$ but the 1.18\,km\,s$^{-1}$ feature had disappeared. This variability is consistent with the findings of \citet{claussen1996} that water maser features associated with low-mass YSOs can have lifetimes of $\lesssim$2 months.

\subsection{EXors}
\subsubsection{V512~Per (SVS~13A)} 
\label{sec:res_v512per}
Located in the low-mass star-forming region NGC~1333, the source V512~Per (commonly known as SVS~13 or SVS~13A, e.g.\ \citealt{plunkett2013}) has been the subject of extensive multiwavelength studies, resulting in a complex nomenclature.
The source was discovered during an infrared survey \citep[SVS76~NGC~1333~13A,][]{strom1976}.  An optical outburst was detected in the late 1980's \citep[][]{mauron1991-iauc} and observations by \citet{Eisloeffel1991} confirmed it showed EXor properties. 
The variable name V512~Per was assigned in the 71$^{st}$ Name-List of Variable Stars by \citet{kazarovets1993}, who noted SVS~13 and V512~Per were the same source. 
A radio counterpart of the optical/near-infrared source, named VLA~4, was first detected by \citet{rodriguez1997} and later resolved into a binary \citep[VLA~4A and 4B;][]{anglada2000}. \citet{rodriguez2002} note that SVS~13 (therefore V512~Per) and VLA~4 are the same source, consistent with other studies \citep[see, e.g.,][]{goodrich1986,fujiyoshi2015}. The source is also commonly known as SVS~13A \citep[see, e.g.,][and references therein]{plunkett2013}
and is associated with several Herbig/Haro objects 
\citep[HH~7--11; e.g.,][]{rodriguez1997,bachiller2000}.
In this paper we refer to the source as V512~Per, noting that this name might be more familiar to the variable star community \citep[e.g.,][]{kazarovets1993,audard2014} while SVS~13, VLA~4, or SVS~13A may be more familiar to the radio astronomy community \citep[e.g.,][]{rodriguez2002,plunkett2013}.

\begin{figure}[h!]
\centering 
\includegraphics[width=\columnwidth]{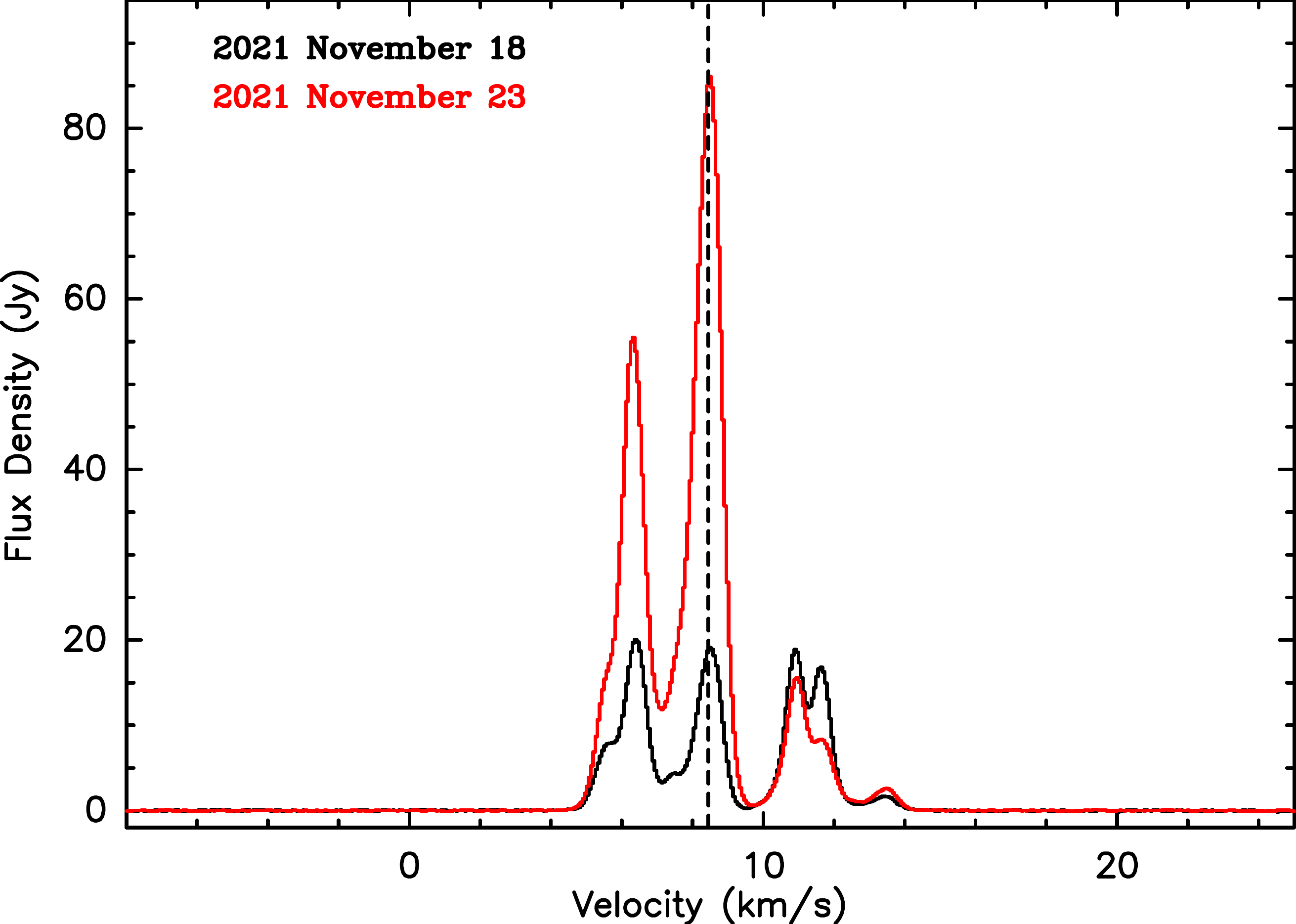}
\caption{H$_2$O maser spectra toward V512~Per in November 2021: the two epochs are indicated at upper left. The dashed line indicates the average $\varv_{\rm LSR}$ of 8.39\,km\,s$^{-1}$ derived from the NH$_3$ (1,1), (2,2) and (3,3) transitions \citep{szabo2023}.}
\label{fig:v512per}
\end{figure}

Figure~\ref{fig:v512per} shows the spectra obtained towards V512~Per in 2021 November.
On 2021 November 18, we detected at least 6 maser features towards V512~Per (see Figure~\ref{fig:v512per}), with the brightest one being 20.2\,Jy. Here we note that only 5 of them are shown in Table~\ref{tab:fit_results}, since the additional feature on the main component at 6.31\,km\,s$^{-1}$ cannot result in a reliable Gaussian fit.  Five days later, two velocity features, at 6.3\,km\,s$^{-1}$ and 8.4\,km\,s$^{-1}$, had increased in flux density by factors of $\sim$3 and $\sim$4, respectively.
Previous H$_2$O maser observations of the V512~Per region revealed three maser positions, H$_2$O(A), H$_2$O(B), and H$_2$O(C) \citep{haschick1980}. H$_2$O(A) is associated with V512~Per. H$_2$O(B), also known as HH~7-11(B), VLA~2, SVS~13C, or MMS3, is a Class\,{\footnotesize 0} source located $\sim$0.5$'$, to the southwest \citep{cesaroni1988,segura-cox2018,chen2013, plunkett2013}, while H$_2$O(C) is $\sim$2.5$'$ southeast of V512~Per \citep[][]{haschick1980}. 

To investigate which of the observed velocity components may be associated with V512~Per, 
we carried out a nine-point grid of observations centred on V512~Per on 2022 February 5 (with pointings separated by 20\arcsec). 
The results indicate that the strong water maser features at  5--10~km\,s$^{-1}$ are brightest at an offset position ($-$20\arcsec,$-$20\arcsec) rather than toward V512~Per (0\arcsec,0\arcsec), suggesting that these maser features do not arise mainly from V512~Per. The $\sim$12\,km\,s$^{-1}$ component, in contrast, is strongest towards V512~Per and is likely associated with the eruptive source (Figure~\ref{fig:svs13-vs-h2ob}, see also Figure~\ref{fig:map}).

\subsection{Serendipitous detections towards Class\,{\footnotesize 0} protostars}
\subsubsection{A water maser flare in H$_2$O(B)}
\label{sect:h2ob-followup}

\begin{figure}[!htbp]
\centering 
\includegraphics[width=1\columnwidth]{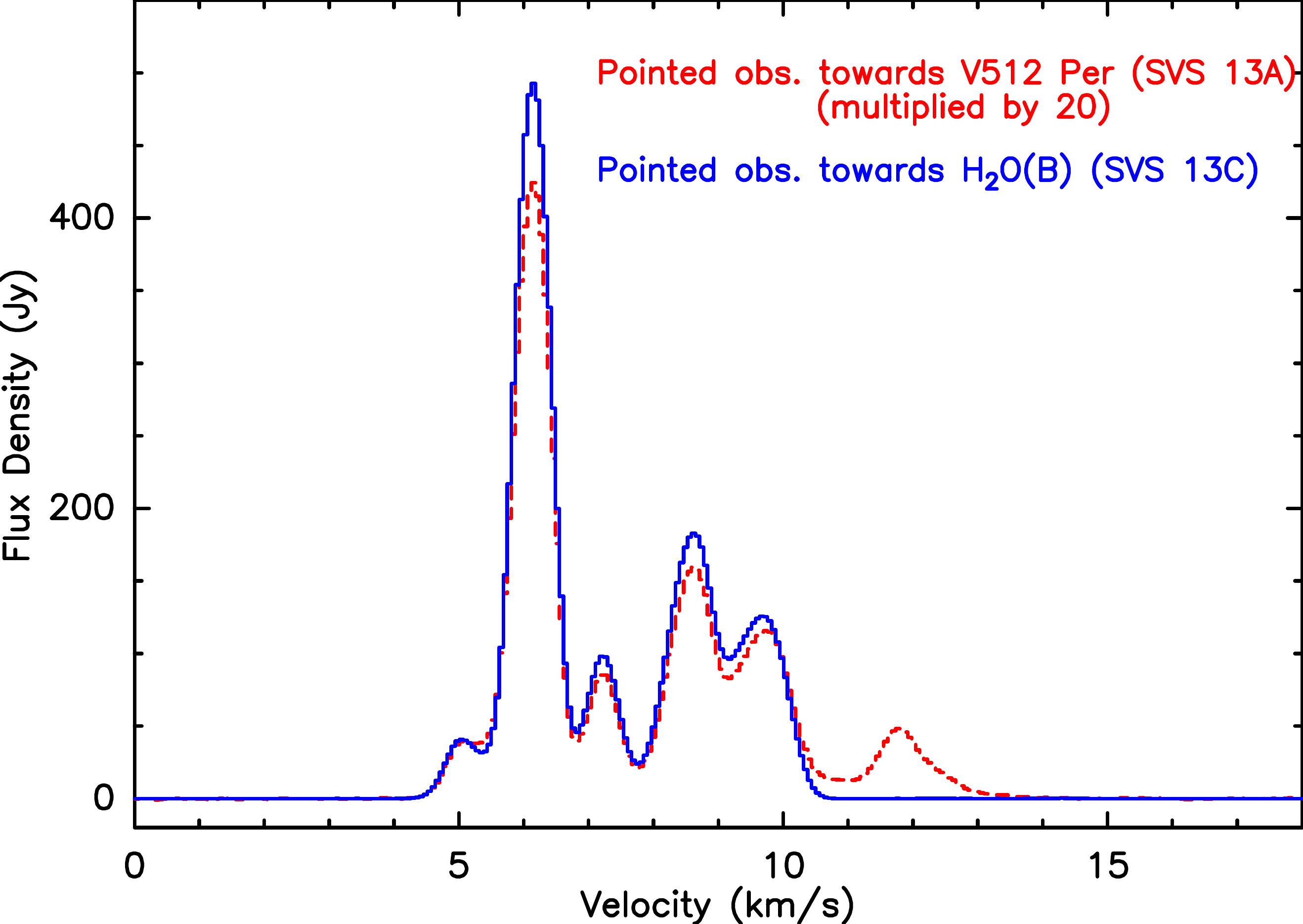}
\caption{Pointed H$_2$O maser spectra towards V512~Per and H$_2$O(B) observed on 2022 February 5. The spectrum of V512~Per is multiplied by 20 to better match the spectrum of H$_2$O(B).}
\label{fig:svs13-vs-h2ob}
\end{figure}

\begin{figure*}[!htbp]
\centering 
\includegraphics[width=0.95\textwidth]{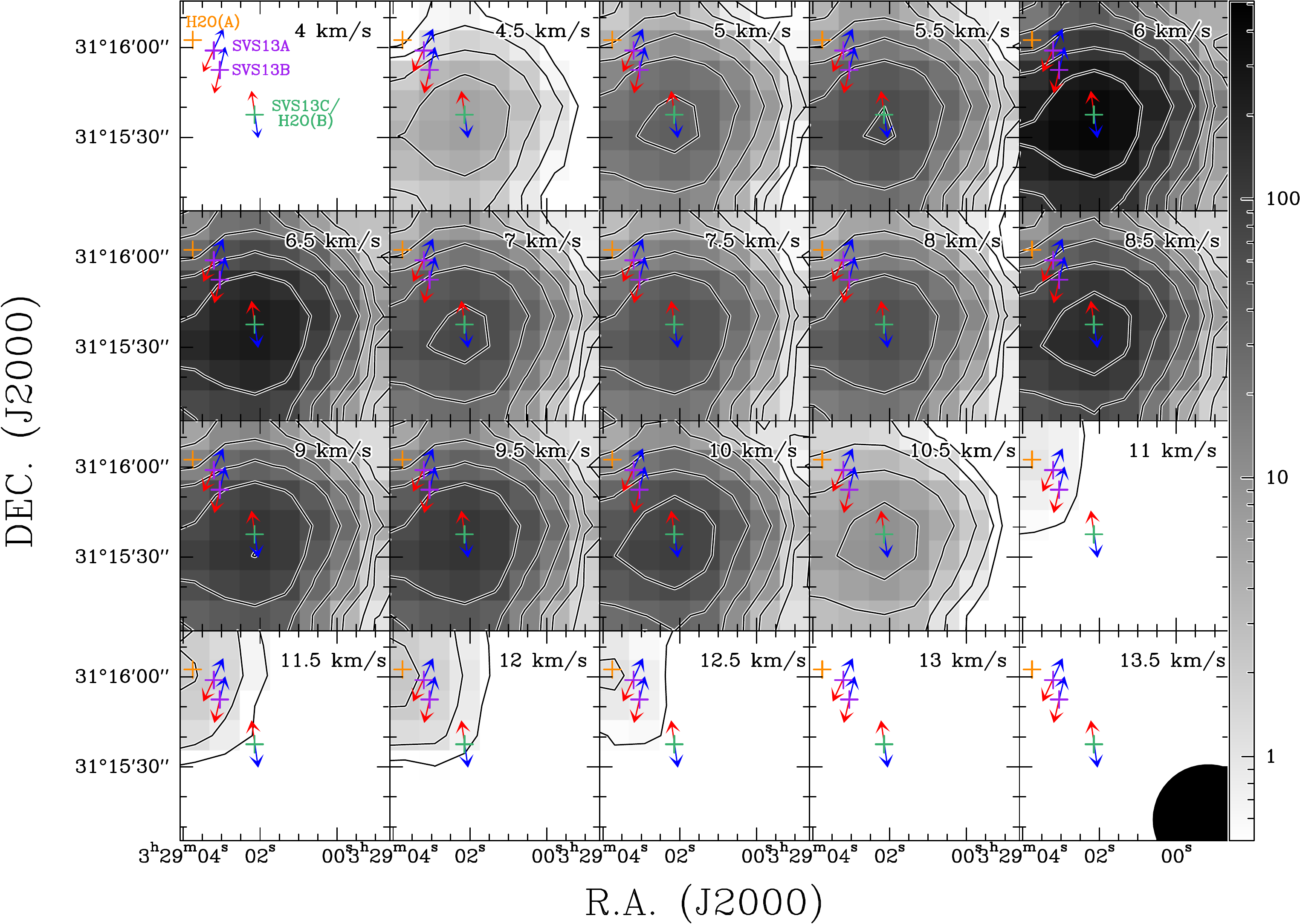}
\caption{Channel maps of H$_{2}$O masers in H$_2$O(B) (SVS~13C) and V512~Per (SVS~13A).
The contours start at 0.5\,Jy, and then increase by a factor of two. The plus signs represent the positions of the two H$_{2}$O masers (orange and green) previously detected by \citet{haschick1980} and YSOs \citep[purple; e.g.,][]{plunkett2013}. 
Based on previous observations \citep{plunkett2013,2021A&A...648A..45P}, the outflow directions are indicated by red and blue arrows. The beam size is shown in the lower right corner of the last panel. The colour bar represents the flux density in units of Jy.}
\label{fig:map}
\end{figure*}

In addition to the nine-point map described above (Sect.~\ref{sec:res_v512per}), we also performed OTF mapping towards V512~Per and H$_2$O(B), shown in Figure~\ref{fig:map}.
As illustrated by the channel maps in Figure~\ref{fig:map}, spectral features at $\varv_{\rm LSR}\leq$11\,km\,s$^{-1}$ peak around H$_{2}$O(B) while spectral features at $\varv_{\rm LSR}>$11\,km\,s$^{-1}$ peak around V512~Per. 
Figure~\ref{fig:svs13-vs-h2ob} compares our pointed observations toward V512~Per and H$_{2}$O(B) on 2022 February 5: the spectra show very similar profiles between 4\,km\,s$^{-1}$ and $\sim$10\,km\,s$^{-1}$ but the intensities are different by a factor of $\sim$20. 
This similarity suggests that our pointed observations of V512~Per, including those shown in Figure~\ref{fig:v512per}, have significant contributions from H$_{2}$O(B). 
We estimate this contribution for our 2022 February 5 observations assuming a perfect Gaussian beam pattern with a beam size of 40$\arcsec$. A source at an offset of 38.7$\arcsec$ (the angular separation between V512~Per and H$_2$O(B) derived from our observations, see Table~\ref{tab:coordinates}) will fall at the 7.5\% response level of the beam, or between the 3.7--14\% levels assuming a typical pointing error of 5$\arcsec$. 
Thus H$_2$O(B), with a flux density of 498.7\,Jy, would contribute 18.4--69.8\,Jy to the spectrum observed towards V512~Per, comparable to the observed value of 21.3\,Jy (Table~\ref{tab:fit_results}).

Notably, in our pointed 2022 February 5 observations, the peak flux density of the water maser in H$_{2}$O(B) is 498.7\,Jy at $\varv_{\rm LSR}=$ 6.1\,km\,s$^{-1}$.
This is the highest flux density reported for this source to-date \citep[c.f.][]{haschick1980,lyo2014}, indicative of a maser flare (see also Sect.~\ref{sec:long-term}).

\subsubsection{RNO~1B/1C and IRAS~00338$+$6312}
\label{sec:fuors-rno1b1c}

RNO~1B/1C (V710~Cas) is a double FUor system, with both RNO~1B and 1C classified as FUors \citep{staude1991,kenyon1993}. The binary is part of a cluster of deeply embedded YSOs \citep[e.g.,][]{quanz2007} that has been targeted by numerous water maser studies
\citep[e.g.,][and references therein]{fiebig1995,furuya2003,sunada2007,bae2011}.
Previous VLA observations of the clustered region suggest that the water masers between $\varv_{\rm LSR}\sim-$30\,km\,s$^{-1}$ and $\varv_{\rm LSR}\sim-$5\,km\,s$^{-1}$ originate from the deeply embedded Class\,{\footnotesize 0} object IRAS~00338$+$6312 rather than RNO~1B or RNO~1C \citep[see, e.g.,][and references therein]{fiebig1995,1996A&A...310..199F}.
IRAS~00338$+$6312 is only $\sim$4\arcsec\,northeast of RNO~1C, but is a separate object \citep[e.g.,][]{mookerjea1999,quanz2007}. The blueshifted and redshifted masers are thought to arise from the bipolar outflow or an accretion disk \citep{1996A&A...310..199F}.

\begin{figure}[!htbp]
\includegraphics[width=\columnwidth]{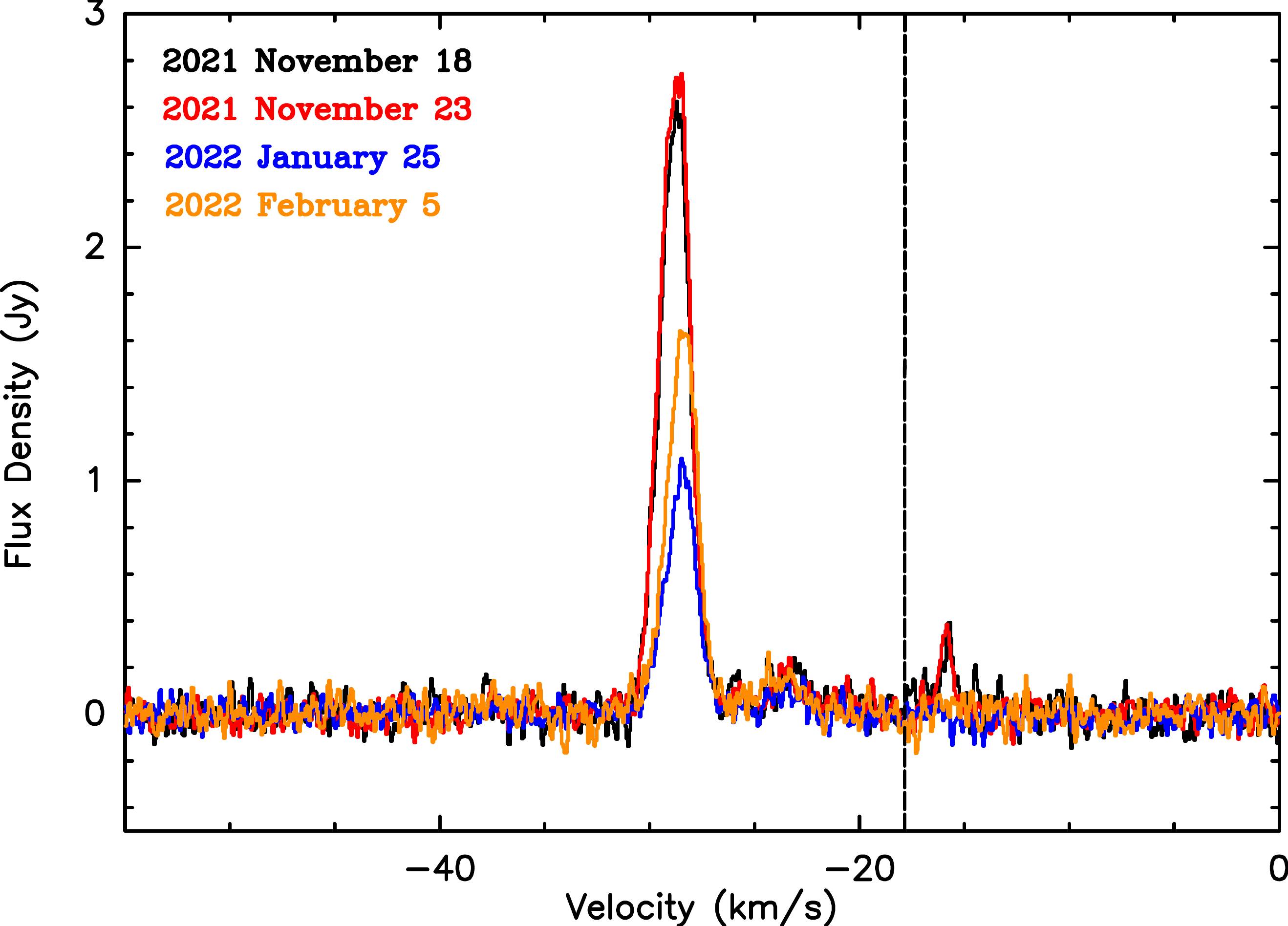}
\caption{H$_2$O maser observed in IRAS~00338$+$6312 at four epochs in 2021 November, 2022 January and February. The dashed vertical line indicates the systemic $\varv_{\rm LSR}$ of $-$17.83\,km\,s$^{-1}$ derived from the NH$_3$ (1,1), (2,2) and (3,3) transitions \citep{szabo2023}.}
\label{fig:rno1b1c}
\end{figure}

In our pointing towards RNO 1B/1C, we detected water maser emission in four epochs, as shown in Figure~\ref{fig:rno1b1c}.
During our first observations on 2021 November 18, we detected two maser features at $\varv_{\rm LSR}=-$28.78\,km\,s$^{-1}$ and $\varv_{\rm LSR}=-$15.79\,km\,s$^{-1}$, and five days later the flux densities and LSR velocities of the two maser features were nearly unchanged. The source was observed again on 2022 January 25 and February 5: in these observations, the $\varv_{\rm LSR}\sim -$15.8\,km\,s$^{-1}$ feature had disappeared and the blueshifted maser was weaker and had slightly shifted in velocity, to  $\varv_{\rm LSR}\sim -$28.48\,km\,s$^{-1}$. 
The 3$\sigma$ upper limits for the $\varv_{\rm LSR}\sim -$15.8\,km\,s$^{-1}$ feature are 0.12\,Jy and 0.15\,Jy for the observations on 2022 January 25 and February 5, respectively. 
We also note that the $\varv_{\rm LSR}\sim-$28\,km\,s$^{-1}$ feature has the largest velocity offset with respect to the cloud among our detections, $\sim$10\,km\,s$^{-1}$ (see Table~\ref{tab:fit_results}). 
Based on comparing our results to the literature, 
the water maser features detected in our survey are most likely to originate from IRAS~00338$+$6312 rather than RNO~1B/1C. The velocities of our detected masers are similar to those of the masers associated with IRAS 00338$+$6312 in the VLA observations \citep{fiebig1995,1996A&A...310..199F}, and also match the velocity range of the molecular outflow \citep[about $-$30\,km\,s$^{-1}$ to $-$5\,km\,s$^{-1}$][]{snell1990,yang1991} driven by IRAS 00338+6312 \citep{henning1992,wouterloot1993,anglada1994,furuya2003,bae2011}.
We therefore do not count the water maser emission in our RNO~1B/1C pointing as a detection towards an eruptive star, and the 3$\sigma$ upper limits are given in Table~\ref{tab:appendix_non-detections}.

\section{Discussion} \label{discussion}
\subsection{Long-term time variation}\label{sec:long-term}
Water maser flares have been  recognized in star forming regions for decades \citep[e.g.,][]{1998ApJ...509..256B,2018IAUS..336..279K}, with recent observations suggesting that water maser flares can accompany ejection events associated with accretion bursts in massive and intermediate-mass stars \citep[e.g.,][]{2018MNRAS.478.1077M,brogan2018,2021ApJ...922...90C,2022A&A...664A..44B}. Hence, one might expect such water maser flares from FUors/EXors. We therefore investigate if our targets have experienced water maser flares.

\begin{figure}[htbp!]
\centering
\begin{minipage}[b]{9cm}
\includegraphics[width=\columnwidth]{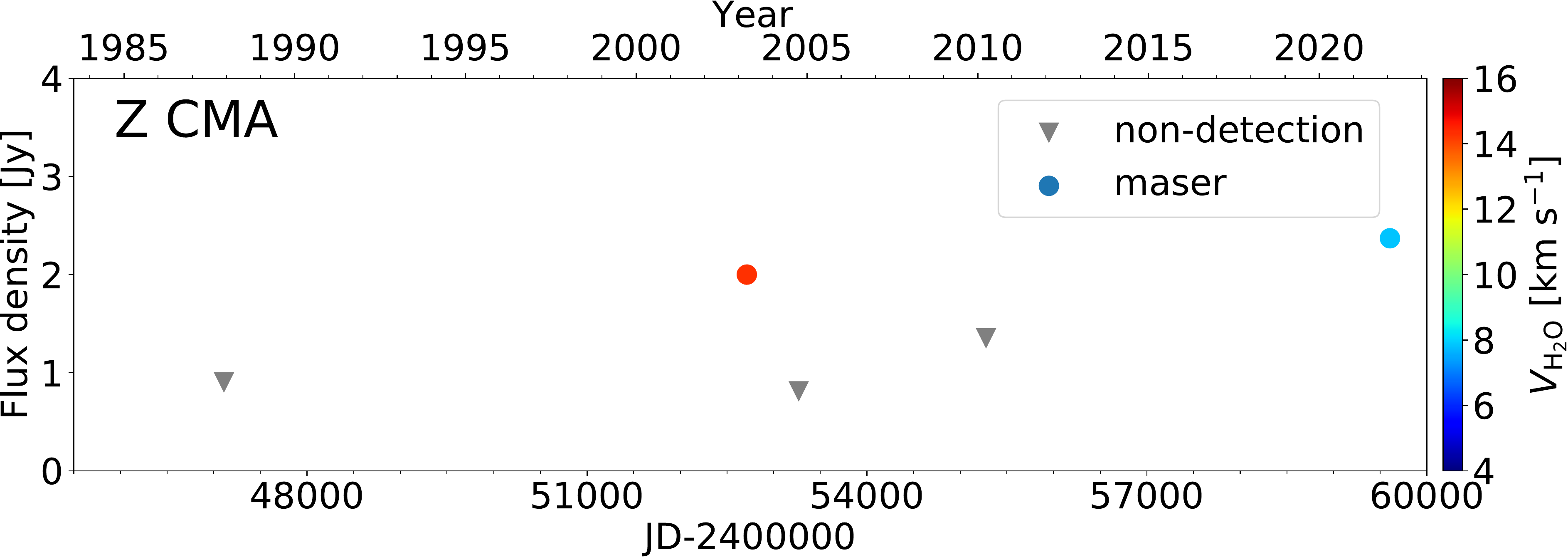}
\vspace{-2mm}
\end{minipage}
\begin{minipage}[b]{9cm}
\includegraphics[width=\columnwidth]{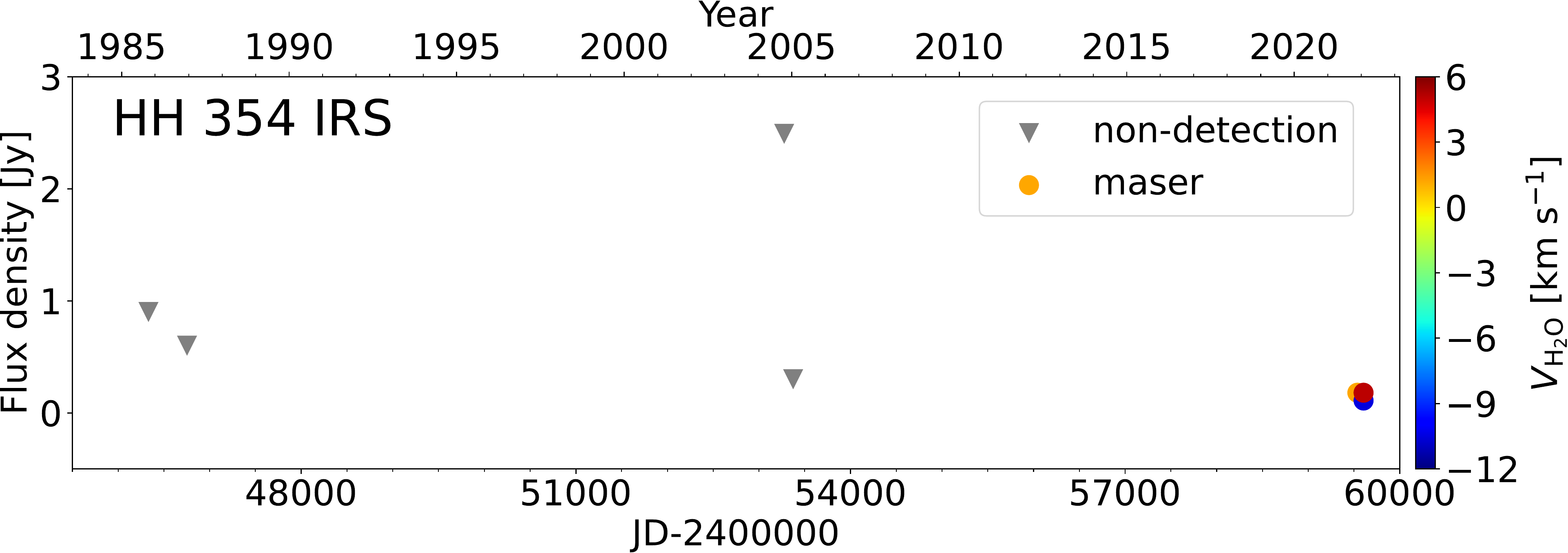}
\vspace{-3mm}
\end{minipage}
\begin{minipage}[b]{9cm}
\includegraphics[width=\columnwidth]{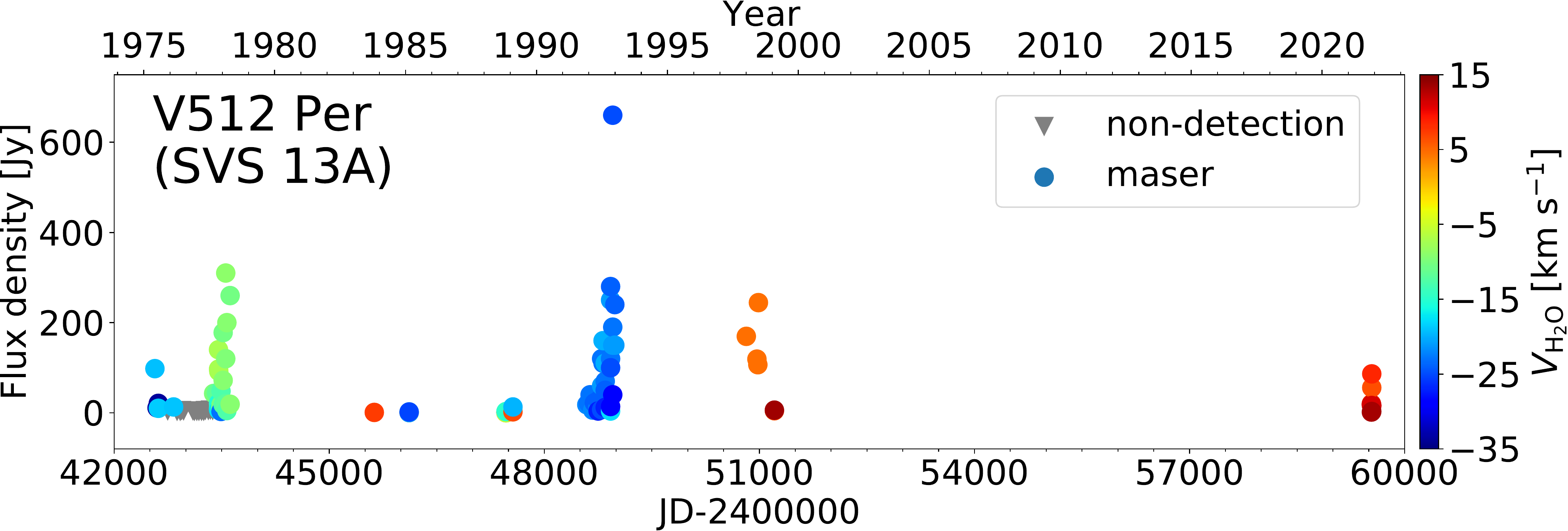}
\vspace{-2mm}
\end{minipage}
\begin{minipage}[b]{9cm}
\includegraphics[width=\columnwidth]{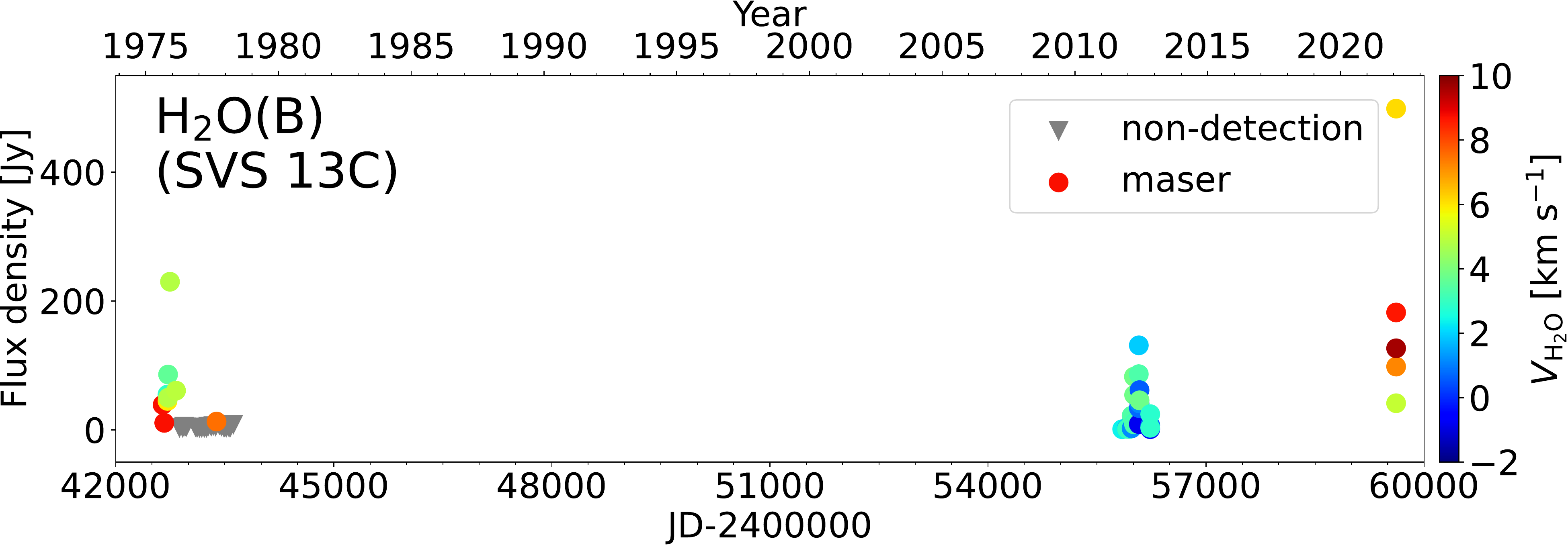}
\vspace{-2mm}
\end{minipage}
\begin{minipage}[b]{9cm}
\includegraphics[width=\columnwidth]{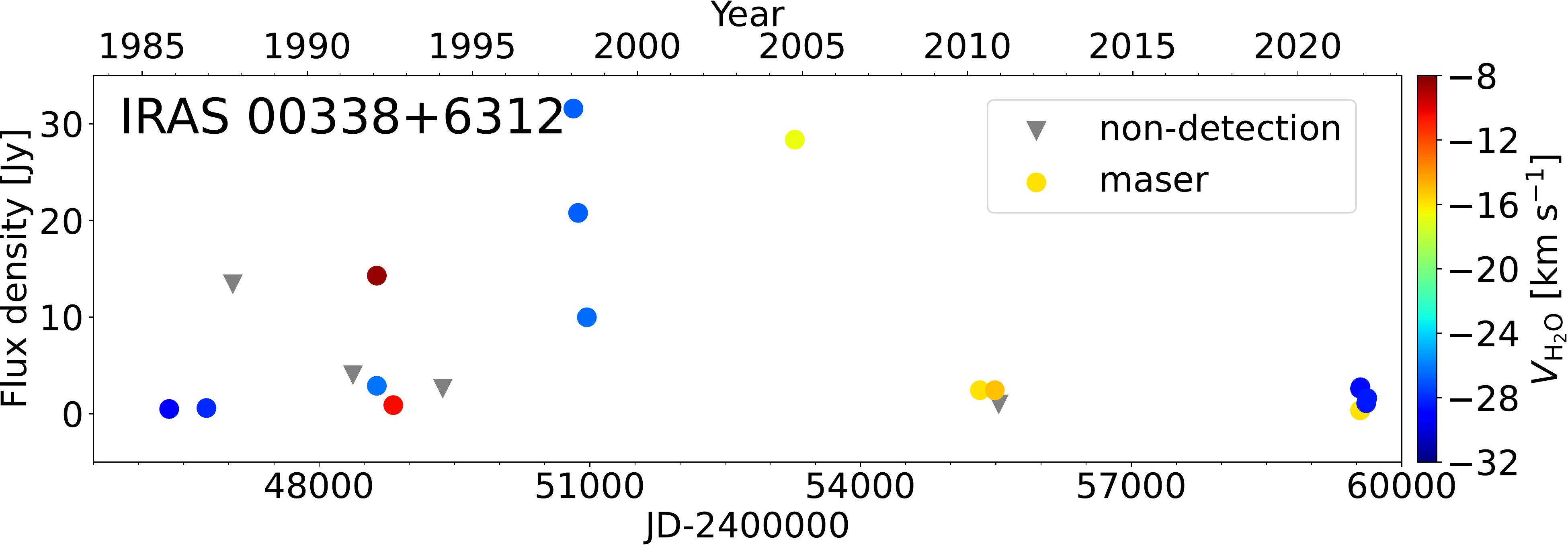}
\vspace{-2mm}
\end{minipage}
\caption{Long-term variations in flux density of detected water masers. The colour-coded dots and grey triangles represent flux densities for maser detections, colour-coded by velocity (see colour-bar at right), and upper limits for non-detections, respectively. References for archival data are given in Sect.~\ref{sec:long-term}.}
\label{fig:long-term}
\end{figure}

Figure~\ref{fig:long-term} presents long-term time series for the water masers detected in our survey, which show that these masers are quite variable in both flux density and LSR velocity. 
Based on data from the literature, Z~CMa appears to be in a relatively active phase, with the flux density of 2.4~Jy during our observations the highest observed to date  \citep[c.f.][]{blitz1979,thum1981,deguchi1989,scappini1991,palla1993,moscadelli2006,sunada2007,bae2011,kim2018}. 
For HH~354~IRS, no water maser emission was detected by previous observations \citep{wouterloot1993,persi1994,sunada2007}. We report the first water maser detection toward this source. Since the upper limits of previous observations are comparable to the detected flux densities (see Figure~\ref{fig:long-term}), we cannot conclude whether the maser was in its active or quiescent phase during our observations.
For V512~Per, Figure~\ref{fig:long-term} compares the velocity component in our observations that likely arises from V512~Per (see Sect.~\ref{sec:res_v512per}) to archival data that include both single-dish and interferometric measurements \citep{haschick1980,claussen1996,rodriguez2002,furuya2003}.We note that in the case of the \citet{claussen1996} data, the results were measured from the published figures.
Based on this comparison, we identify three water maser flares, in 1978, 1992, and 1998, which reached peak 
flux densities of $\sim$310\,Jy, 660\,Jy, and 244\,Jy on 1978 February 17, 1992 November 28, and 1998 June 22, respectively.  The observations spanning these dates were performed with single-dish telescopes with large beams ($>$1\arcmin), so H$_2$O(B) could potentially contribute to the observed flux densities (see Sect.~\ref{sec:res_v512per}). \citet{claussen1996} note, however, that the maser features detected in their 1991-92 observations all had velocities consistent with those of H$_2$O(A)/V512~Per, suggesting that this flare was associated with the eruptive star.    

For H$_{2}$O(B), we find no suggestion in the literature of this source being an eruptive variable at optical or near-infrared wavelengths, but our comparison with previous water maser observations \citep[Figure~\ref{fig:long-term};][]{haschick1980,lyo2014} shows three maser flares with peak flux densities of $>$100\,Jy, on 1975 November 30, 2012 May 28, and 2022 February 5. 
As for V512~Per, Figure~\ref{fig:long-term} compares
the velocity components in our observations that likely arise from H$_2$O(B) (Sect.~\ref{sec:res_v512per}\&\ref{sect:h2ob-followup}) with historical data.
Again, the large single dish beams encompass both H$_2$O(B) and V512~Per, meaning that we cannot rule out a contribution from V512~Per to the historical flares. For instance, the observations of \citet{lyo2014} had a HPBW of 120$\arcsec$. As noted in Sect.~\ref{sect:h2ob-followup}, the water maser flare detected in our observations on 2022 February 5 is the brightest to date, with a peak flux density of 498.7\,Jy.

For IRAS~00338$+$6312, there is similarly no suggestion in the literature of this being an eruptive source in the optical or near-infrared, but
Figure~\ref{fig:long-term} suggests its water maser emission 
was in an active phase in 1998 and 2004 \citep{cesaroni1988,henning1992,wouterloot1993,persi1994,fiebig1995,codella1995,furuya2003,sunada2007,bae2011}, but relatively quiescent during our observations. The highest flux density reached was $\sim$31\,Jy on 1998 January 5 \citep{furuya2003}. 

Periodic variations have been reported in some velocity components of the 22.2\,GHz H$_2$O (and the 6.7\,GHz Class\,{\footnotesize II} CH$_3$OH) masers associated with the intermediate-mass YSO G107.298$+$5.639, and cyclic accretion instabilities have been invoked to explain this peculiar behavior \citep{2016MNRAS.459L..56S}. Low-mass stars like FUors and EXors might also experience cyclic accretion events, but we do not find evidence for periodic variations in Figure~\ref{fig:long-term}. 

\subsection{Scarcity of water masers in selected eruptive systems}
\label{sec:detection_rate}
Our water maser detection rate of 6\% in FUors and EXors is perhaps surprising in light of the close connection between water maser emission and mass accretion and ejection in protostars (see Sect.~\ref{sec:intro}). 
In this section, we consider possible explanations for the low detection rate. 

First, the low detection rate could be caused by an evolutionary effect. Previous observations indicate that the water maser detection rate decreases from Class\,{\footnotesize 0} to Class\,{\footnotesize II} objects \citep[e.g.,][]{furuya2001}. Since the selected FUors/EXors are mainly Class\,{\footnotesize I} and Class\,{\footnotesize II} objects (see Tables~\ref{tab:coordinates} and \ref{tab:appendix_non-detections}), one would expect a lower detection rate compared to  Class\,{\footnotesize 0} objects. Furthermore, our detection rate is comparable to that (6.3\%) for Class\,{\footnotesize I} objects in \citet{furuya2001}. We do not detect any water masers toward Class\,{\footnotesize II} objects, which further supports the evolutionary trend proposed by \citet{furuya2001}.

Second, water masers have relatively low luminosities in low-mass star formation regions. Statistical studies have shown that the maser luminosities are correlated with bolometric luminosities \citep[e.g., Figure~16 in][]{urquhart2011}. This suggests lower maser luminosities in low-mass star formation regions, so lower flux densities would be expected. This could contribute to our low detection rate toward low-mass eruptive stars. This is supported by previous water maser surveys toward the Serpens South and Orion molecular clouds \citep{kang2013,2021AJ....162...68O}, which give detection rates of $\lesssim$2\% for low-mass protostars.

Third, water masers show rapid time variations. The time variability of water masers is evident in our study (see also Figures~\ref{fig:hh354}, \ref{fig:svs13-vs-h2ob} and \ref{fig:long-term}). Water masers can be in a quiescent phase for $\sim$5\,years \citep{claussen1996}, meaning that maser emission would not be detected during that time even for sources known to be associated with water masers. This is consistent with the fact that several water masers reported by previous studies are not detected in our observations (see Table~\ref{tab:appendix_non-detections}). It is possible that non-detection of water masers is due to their inactive state. 
Indeed, including historical detections, the detection rate of water masers in eruptive stars in our sample is $\sim$15\% (excluding the unclassified Gaia alerts), which is higher than our survey detection rate of 6\%, suggesting that previously detected water masers were in an inactive phase during our observations.

\section{Conclusions} 
\label{sec:conclusions}
In this paper, we presented the results of the first dedicated water maser survey towards FUors and EXors, two classes of low-mass young eruptive stars. We detected H$_2$O masers toward five objects, of which three are young eruptive stars: Z~CMa (FUor; Class\,{\footnotesize I}), HH~354~IRS (FUor; Class\,{\footnotesize 0/I}), V512~Per (EXor; Class\,{\footnotesize I}), IRAS~00338$+$6312 (Class\,{\footnotesize 0}) and H$_2$O(B) (Class\,{\footnotesize 0}).
Our detection is the first report of water maser emission in HH~354~IRS.
Our observations reveal the highest peak flux density yet reported towards H$_2$O(B) (498.7\,Jy), indicative of a recent H$_2$O maser flare. Overall, our observations result in a detection rate of $\sim$6\% for young eruptive stars.
Analysis of the long-term time series of the water masers suggests that V512~Per and H$_{2}$O(B) have experienced multiple water maser flares.

Despite the low detection rate, our observations have confirmed the presence of 22.2\,GHz water maser emission in FUors and EXors, meaning that follow-up radio interferometric observations can be used to probe the environments of eruptive stars on small scales \citep[see, e.g.,][]{haschick1980,rodriguez2002}.
If water masers are in general weak in FUors/EXors (Sect.~\ref{sec:detection_rate}), deeper observations would also be expected to find more of them.
Expanding on optical and near-infrared knowledge of FUors/EXors with more radio observations, especially future VLBI measurements, will be crucial to better understand the underlying physics (e.g., mass accretion and ejection) of such peculiar objects, and eventually the formation of Sun-like stars. 

\begin{acknowledgements}
We thank the referee for their valuable comments and suggestions which improved the quality of the manuscript.
Based on observations (Project ID: 95-21, PI: Szab\'o) with the 100-m telescope of the MPIfR (Max-Planck-Institut für Radioastronomie) in Effelsberg.
Zs.M.Sz. acknowledges funding from a St Leonards scholarship from the University of St Andrews.
For the purpose of open access, the author has applied a Creative Commons
Attribution (CC BY) licence to any Author Accepted Manuscript version arising.
This project has received funding from the European Research Council (ERC) under the European Union's Horizon 2020 research and innovation programme under grant agreement No 716155 (SACCRED). This work has made use of the database for astrophysical masers\footnote{\url{https://maserdb.net/}} \citep{2022AJ....163..124L}. 
O.B. acknowledges financial support from the Italian Ministry of University and Research - Project Proposal CIR01$\_$00010.
We acknowledge ESA Gaia, DPAC and the Photometric Science Alerts Team (http://gsaweb.ast.cam.ac.uk/alerts).
This work presents results from the European Space Agency (ESA) space mission Gaia. Gaia data are being processed by the Gaia Data Processing and Analysis Consortium (DPAC). Funding for the DPAC is provided by national institutions, in particular the institutions participating in the Gaia MultiLateral Agreement (MLA). The Gaia mission website is https://www.cosmos.esa.int/gaia. The Gaia archive website is https://archives.esac.esa.int/gaia.
\end{acknowledgements}

\bibliographystyle{aa}
\bibliography{paper}
\begin{appendix} 
\section{W75N} \label{sec:w75n}
\begin{figure}[h!]
\centering 
\includegraphics[width=\columnwidth]{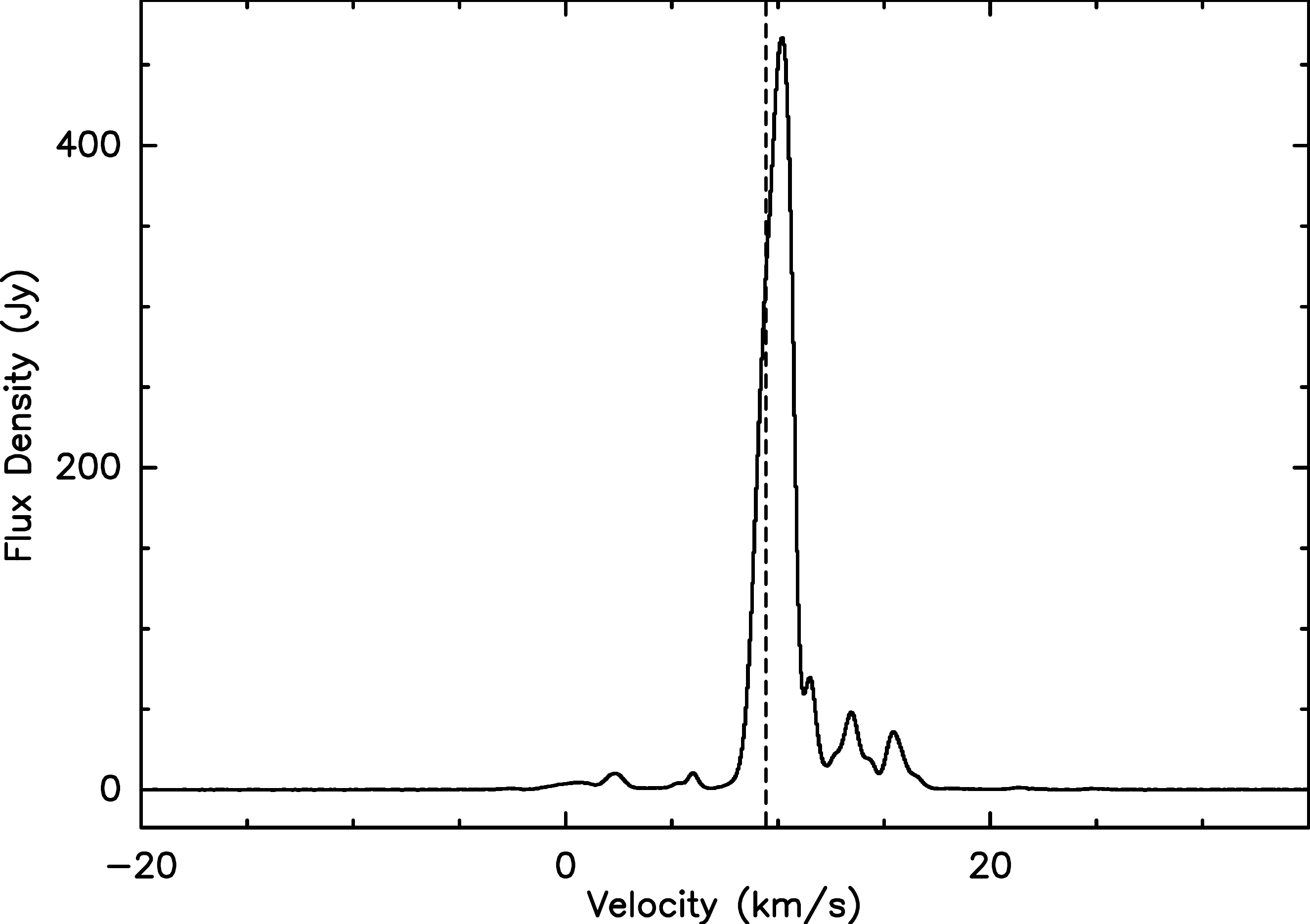}
\caption{H$_2$O maser spectrum observed toward W75N on 2021 November 18. The dashed line indicates the centroid LSR velocity of 9.43\,km\,s$^{-1}$ from the NH$_3$ (1,1), (2,2) and (3,3) transitions.}
\label{fig:w75n}
\end{figure}

During our initial observing run on 2021 November 18, our setup was verified by observing W75N, a well-known massive star-forming region showing bright water masers \citep[][and references therein]{lekht1984,hunter1994,jeong-sook2013}. Although our survey focused on low- and intermediate-mass young stars, here we briefly present the W75N maser spectrum to make it available for potential future studies of maser variability in this region. 
Figure~\ref{fig:w75n} shows the water maser spectrum observed toward W75N on 2021 November 18. The dashed vertical line indicates the average centroid  $\varv_{\rm LSR}$ of 9.43\,km\,s$^{-1}$ derived from the NH$_3$ (1,1), (2,2) and (3,3) transitions. In Table~\ref{tab:w75n}, we list the properties of the water maser features, along with the  $\varv_{\rm LSR}$ results from the NH$_3$ transitions.

\begin{table}[!htbp]
\tiny
\caption{H$_2$O maser velocity components and NH$_3$ LSR velocities from observations of W75N on 2021 November 18.}
\label{tab:w75n}      
\centering                                      
\begin{tabular}{rrrrrrrr}    
\hline\hline          
\multicolumn{4}{c}{H$_2$O}   & \multicolumn{1}{c}{NH$_3$ (1,1)} & \multicolumn{1}{c}{(2,2)} & \multicolumn{1}{c}{(3,3)} \\
\cline{1-4} 
\cdashline{5-5}
\cdashline{6-6}
\cdashline{7-7}
\multicolumn{1}{c}{$\varv_{\rm LSR}$} & \multicolumn{1}{c}{RMS} & \multicolumn{1}{c}{$S_{\nu}^*$} & \multicolumn{1}{c}{$L_{\rm H_2O}$} & \multicolumn{1}{c}{$\varv_{\rm LSR}$} & \multicolumn{1}{c}{$\varv_{\rm LSR}$}  & \multicolumn{1}{c}{$\varv_{\rm LSR}$}    \\
\multicolumn{1}{c}{(km s$^{-1}$)} & \multicolumn{1}{c}{(Jy)} & \multicolumn{1}{c}{(Jy)}  & \multicolumn{1}{c}{(\(\textup{L}_\odot\))} & \multicolumn{1}{c}{(km s$^{-1}$)} & \multicolumn{1}{c}{(km s$^{-1}$)} & \multicolumn{1}{c}{(km s$^{-1}$)}  \\
\hline \hline                                 
$0.84$ $(0.01)$   & $0.08$  & $11.27$  & 3.06$\times10^{-6}$ & \multirow{5}{*}{$9.48$ $(0.03)$} & \multirow{5}{*}{$9.37$ $(0.04)$} & \multirow{5}{*}{$9.46$ $(0.07$)}\\
$5.88$ $(1.23)$   & $0.08$  & $11.27$  & 3.91$\times10^{-7}$ \\
$9.99$ $(0.03)$   & $0.08$  & $467.64$  & 3.23$\times10^{-5}$ \\
$13.35$ $(0.06)$   & $0.08$  & $49.71$  & 3.11$\times10^{-6}$  \\
$15.56$ $(0.06)$   & $0.08$  & $36.50$  & 1.32$\times10^{-6}$  \\
\hline \hline
\end{tabular}
\tablefoot{$^*$Results from the Gaussian fitting, for more information, see Sect.~\ref{sec:observations}.}
\end{table}

\section{Non-detections in our survey} \label{sec:app_non-detections}
In total, our survey consisted of 51 objects: 33 Fuors, 13 EXors and 5 Gaia alerts accessible with the Effelsberg\,100-m telescope. The chosen Gaia~alerts are part of the Piszkéstető Monitoring Program, in Hungary, which started a few years ago with the aim of following the optical wavelength brightness variations of Gaia alert sources with light curves that resemble those of FUors and EXors
\citep[see e.g.,][]{szegedi-elek2020,nagy2021,cruz-saenz2022}.
We chose Gaia alerts for inclusion in our sample based on their having light curves and luminosities similar to those of FUors and EXors. 
Water masers were detected in only four out of our 51 targeted sources, however the emission in one of those target pointings is not attributed to the targeted eruptive star (see Sect.~\ref{sec:fuors-rno1b1c}). The 48 non-detections, within an expected velocity range from $-$100 to $+$100\,km\,s$^{-1}$, are reported in Table~\ref{tab:appendix_non-detections}.

\begin{table*}[!htbp]
\tiny
\caption{H$_2$O maser non-detections in our survey.}
\centering
\label{tab:appendix_non-detections}      
\begin{tabular}{cccccccccc}        
\hline\hline                        
\multirow{3}{*}{Source} &  Type         & R.A. (J2000)        & Dec. (J2000) & Upper limit$^{(a)}$      & Previous survey$^{(b)}$ & Date$^{(c)}$ & Classification & \multirow{3}{*}{Ref.} & D$^{(d)}$ \\ 
                        & (FUor/EXor/   & \multirow{2}{*}{($^{\rm h}$ $^{\rm m}$ $^{\rm s}$)}   & \multirow{2}{*}{($^{\circ}$ $\arcmin$ $\arcsec$)}   &  \multirow{2}{*}{(Jy)} & (No/Yes/       & \multirow{2}{*}{(A/B/C)} & \multirow{2}{*}{(Class\,{\footnotesize 0 -- II})} & & \multirow{2}{*}{(pc)} \\
                        & Gaia alert)   &                      &   & & Unknown)       & \\
\hline \hline      
V1180~Cas               & EXor   & 02 33 01.53 & +72 43 26.8                & $0.17$ &  Unk. & \multirow{34}{*}{A} & $-$ & $-$ & 908 [51]\\
PP~13S                  & FUor   & 04 10 41.09 & +38 07 54.5                & $0.18$ &  Yes [1] & &  \,{\footnotesize I} &   21 & 350 \\ 
L1551~IRS~5             & FUor   & 04 31 34.07 & +18 08 04.9                & $0.17$ &  No [2,3] & &  \,{\footnotesize I} &  22 & $-$ \\  
XZ~Tau                  & EXor   & 04 31 40.08 & +18 13 56.6	                & $0.18$ &  Unk. & &  \,{\footnotesize II} & 23 & 140 \\ 
UZ~Tau~E                & EXor   & 04 32 43.02 & +25 52 30.9                & $0.18$ &  Unk. & &  \,{\footnotesize II} & 24 & 130 [51] \\
LDN~1415~IRS            & EXor   & 04 41 37.50 & +54 19 22.0                 & $0.21$ &  Unk. & &  \,{\footnotesize I} & 25 & 170 \\ 
DR~Tau                  & EXor   & 04 47 06.21 & +16 58 42.8                 & $0.17$ &  Unk. & &  \,{\footnotesize II} & 26 & 192 [51] \\
V582~Aur                & FUor   & 05 25 51.97 & +34 52 30.0                & $0.15$ &  Unk. & &  \,{\footnotesize II} & 27 & 2401 [52,53] \\
V1118~Ori               & EXor   & 05 34 44.98 & $-$05 33 41.3                & $0.21$ &  Unk. & &  \,{\footnotesize II} & 28 & 414 \\ 
Haro~5a~IRS             & FUor   & 05 35 26.74 & $-$05 03 55.0              & $0.21$ &  Unk. & &  \,{\footnotesize 0/I} & 29 & 450 \\ 
NY~Ori                  & EXor   & 05 35 36.00 & $-$05 12 25.2                & $0.22$ &  Unk. & & $-$ & $-$ & 403 [51] \\
V1143~Ori               & EXor   & 05 38 03.89 & $-$04 16 42.8                & $0.22$ &  Unk. & &  \,{\footnotesize II} & 30 & 395 [51] \\ 
V883~Ori                & FUor   & 05 38 18.09 & $-$07 02 25.9                 & $0.20$ &  No [4,5]      & &  \,{\footnotesize I} & 31 & 417 [54] \\
HBC~494                 & FUor   & 05 40 27.45 & $-$07 27 30.0                & $0.21$ &  Unk. & &  \,{\footnotesize I} & 32 & $-$ \\  
V2775~Ori               & FUor   & 05 42 48.48 & $-$08 16 34.7                & $0.25$ &  No [6]      & & late  \,{\footnotesize I} & 33 & 420 \\   
FU~Ori                  & FUor   & 05 45 22.37 & +09 04 12.3                & $0.22$ &  Unk. & &  \,{\footnotesize II} & 34 & 407 [51] \\    
V1647~Ori               & FUor   & 05 46 13.13 & $-$00 06 04.8                & $0.23$ &  Unk. & &  \,{\footnotesize I/II} & 35 & 412 [51] \\   
AR~6A/6B                & FUor   & 06 40 59.30 & +09 35 49.0                & $0.23$ &  Unk. & &  \,{\footnotesize II} & 36 & 800 \\  
Parsamian~21            & FUor   & 19 29 00.84 & +09 38 43.4	                & $0.16$ &  Unk. & &  \,{\footnotesize I/II} & 37 & 400 \\ 
I~18270-0153W            & FUor  & 18 29 36.90 & $-$01 51 02.0 &  $0.19$ &  No [7]      &  &  \,{\footnotesize I} & 38 & $-$ \\ 
OO~Ser                  & FUor   & 18 29 49.13 & +01 16 20.6                   & $0.18$ &  No [3,8]      & &  \,{\footnotesize I} & 38 & 311 \\ 
V371~Ser                & EXor   & 18 29 51.21 & +01 16 39.4	                 & $0.18$ &  No [3,8]      & & $-$ & $-$ & 311 \\ 
Gaia21aul               & Gaia alert & 18 30 06.18 & 00 42 33.30             & $0.17$ &  No [9,10]      & & $-$ & $-$ & 378 [55]  \\ 
I~18341-0113S            & FUor  & 18 36 45.70 & $-$01 10 29.0 & $0.19$ &  No [3,7]      & &  \,{\footnotesize I} & 38 & $-$ \\ 
Gaia21aru               & Gaia alert & 19 00 56.41 & 18 48 29.20             & $0.16$ &  Unk. & & $-$ & $-$ & $-$ \\
V1515~Cyg               & FUor   & 20 23 48.01 & +42 12 25.7                & $0.18$ &  Unk. & & $-$ & $-$ & 900 [55] \\
PV~Cep & EXor & 20 45 53.90 & $+$67 57 38.6 & $0.17$ & Yes [5,11] & & $-$ & $-$ & 325 \\      
& & & & & No [10,12,13,14,15] & & & & \\
V2492~Cyg               & EXor   & 20 51 26.23 & +44 05 23.8                 & $0.16$ &  No [16]      & &  \,{\footnotesize I} & 39 & 804 [51] \\
V1057~Cyg & FUor & 20 58 53.73 & $+$44 15 28.4 & $0.17$ &  Yes [11]       & & \,{\footnotesize II} & 40 & 891 [55] \\
& & & & & No [3,10,12,13,14] & & & & \\
HBC~722                 & FUor   & 20 58 17.00 & +43 53 43.0                & $0.17$ &  Unk. & &  \,{\footnotesize II} & 41 & 757 [55]  \\
V2495~Cyg               & FUor   & 21 00 25.24 & +52 30 16.9                & $0.16$ &  No [9]      & &  \,{\footnotesize I/II} & 42 & 800 \\ 
RNO~127                 & FUor   & 21 00 31.80 & +52 29 17.0                & $0.17$ &  Unk. & & $-$ & $-$ & 800 \\ 
CB~230                  & FUor   & 21 17 38.62 & +68 17 34.0                & $0.15$ &  No [3,7,17,18]      & &  \,{\footnotesize 0/I} & 43 & 339 [18] \\ 
V1735~Cyg               & FUor   & 21 47 20.66 & +47 32 03.8                & $0.18$ &  No [1,3,12]      & &   \,{\footnotesize II} & 43 & 663 [55] \\
V733~Cep                & FUor   & 22 53 33.25 & +62 32 23.6                & $0.16$ &  Unk. & &  \,{\footnotesize I/II} & 43 & 724 [55] \\
\hline
VY~Tau                  & EXor  & 04 39 17.42 & +22 47 53.3                 & $0.11$ &  Unk. & \multirow{8}{*}{B} &  \,{\footnotesize II} & 44 & 153 [55]  \\    
Gaia21arx               & Gaia alert & 05 36 24.80 & $-$06 17 30.52            & $0.15$ &  Unk. & & $-$ & $-$ & 361 [55]   \\
NGC~2071                & FUor  & 05 47 09.80 & +00 18 00.0                 & $0.15$ &  Yes [1,3,6] & & $-$ & $-$  & $-$ \\ 
I~06297+1021W            & FUor & 06 32 28.70 & +10 19 0                  & $0.16$ &  Unk. & &  \,{\footnotesize I} & 43 & $-$ \\  
I~06393+0913             & FUor & 06 42 08.13 & +09 10 30.0                  & $0.15$ &  Unk. & &  \,{\footnotesize I} & 43 & $-$ \\ 
Gaia18dvy               & FUor    & 20 05 06.02 & $+$36 29 13.52               & $0.14$ &  Unk. & &  \,{\footnotesize II} & 45 & 1880 [45] \\
Gaia19bpg               & Gaia alert & 21 41 50.43 & $+$51 55 45.48            & $0.12$ &  Unk. & & $-$ & $-$ & $-$ \\      
\hline
V899~Mon                & FUor  & 06 09 19.24 & $-$06 41 55.8                 & $0.10$ &  Unk. & \multirow{5}{*}{C} &  \,{\footnotesize II} & 46 & 809 [51] \\
V900~Mon                & FUor  & 06 57 22.22 & $-$08 23 17.6                 & $0.13$ &  Unk. & &  \,{\footnotesize I} & 47 & 1304 [51] \\
V960~Mon                & FUor  & 06 59 31.58 & $-$04 05 27.7                 & $0.12$ &  Unk. & &  \,{\footnotesize II} & 48 & 2189 [51]  \\
iPTF~15AFQ 	            & FUor  & 07 09 21.39 & $-$10 29 34.5                 & $0.13$ &  Unk. & &  \,{\footnotesize I} & 49 & 1315 [51] \\
Gaia20bdk               & Gaia alert   & 07 10 14.92 & $-$18 27 01.04          & $0.15$ &  Unk. & & $-$ & $-$ & $-$ \\   
\hline
\multirow{4}{*}{RNO~1B/1C} & \multirow{4}{*}{FUor} & \multirow{4}{*}{00 36 46.30} & \multirow{4}{*}{$+$63 28 54.0} & 0.05 & \multirow{4}{*}{No [1,7,8,10,19,20]} & A & \multirow{4}{*}{1B:~\,{\footnotesize 0/I--II},~1C:\,{\footnotesize II}} & \multirow{4}{*}{50} & \multirow{4}{*}{965 [55]} \\
 &  &  &  & 0.04 &  & B & & &  \\
 &  &  &  & 0.04 &  & C & & &  \\
 &  &  &  & 0.05 &  & D & & &  \\
\hline \hline
\end{tabular}
\flushleft
\tablefoot{Source types and coordinates are taken from \citet{audard2014}, except for Gaia alerts, for which positions are taken from the Gaia alerts system. \tablefoottext{a}{Upper limits are 3 $\times$ RMS.}\tablefoottext{b}{Previous surveys: No -- Source was observed, but no H$_2$O maser detected; Yes -- H$_2$O maser detected; Unk.~(Unknown) -- To our knowledge, previously not  searched for water masers}.}
\tablefoottext{c}{Our observations were carried out on: A -- 2021-11-18; B -- 2021-11-23; C -- 2022-01-25.; D -- 2022-02-05.}\tablefoottext{d}{Distance -- the default values are from \cite{audard2014}, but in case of more recent data (i.e.~\textit{Gaia}), the values are then updated with references.} \\
References: 1 -- \citet{wouterloot1993}; 2 -- \citet{claussen1996}; 3 -- \citet{sunada2007}; 4 -- \citet{deguchi1989}; 5 -- \citet{takaba2001}; 6 -- \citet{kang2013}; 7 -- \citet{codella1995}; 8 -- \citet{furuya2003}; 9 -- \citet{yung2013}; 10 -- \citet{bae2011}; 11 -- \citet{cesaroni1988}; 12 -- \citet{felli1992}; 13 -- \citet{palagi1993}, 14 -- \citet{palla1993}; 15 -- \citet{valdettaro2001}; 16 -- \citet{urquhart2011}; 17 -- \citet{gomez2006}; 18 -- \citet{brand2019}; 19 -- \citet{henning1992}; 20 -- \citet{fiebig1995}; 21 -- \citet{sandell2001}; 22 -- \citet{fuller1995}; 23 -- \citet{zapatava2015}; 24 -- \citet{mathieau1996}; 25 -- \citet{stecklum2007}; 26 -- \citet{banzatti2014}; 27 -- \citet{abraham2018}; 28 -- \citet{gianni2016}; 29 -- \citet{kospal2017b}; 30 -- \citet{parsamian2004}; 31 -- \citet{white2019}; 32 -- \citet{ruiz-rodriguez2017}; 33 -- \citet{zurlo2017}; 34 -- \citet{herbig1977}; 35 -- \citet{principe2018}; 36 -- \citet{moriarty2008}; 37 --  \citet{kospal2008}; 38 -- \citet{connelley2010}; 39 -- \citet{hillenbrand2013}; 40 -- \citet{feher2017}; 41 -- \citet{kospal2016}; 42 -- \citet{liu2018}; 43 -- \citet{connelley2018}; 44 -- \citet{herbig1990}; 45 -- \citet{szegedi-elek2020}; 46 -- \citet{park2021}; 47 -- \citet{takami2019}; 48 -- \citet{kospal2015}; 49 -- \citet{miller2015}; 50 -- \citet{quanz2007}; 51 -- \citet{gaia-dr3-2022}; 52 -- \citet{bailerjones1}; 53 --\citet{zsidi2019}; 54 -- \citet{menten2007}; 55 -- \citet{bailerjones-edr3}
\end{table*}

\end{appendix}

\end{document}